\documentclass[fleqn,usenatbib]{mnras}

\usepackage{amsmath, amssymb}

\usepackage{newtxtext,newtxmath}

\usepackage[T1]{fontenc}
\usepackage{cleveref}
\usepackage{graphicx}
\usepackage{dcolumn}
\usepackage{bm}
\usepackage[dvipsnames]{xcolor}
\usepackage[ruled,lined]{algorithm2e}
\usepackage{algpseudocode}
\usepackage{caption}
\usepackage{subcaption}
\usepackage{float}

\definecolor{blue}{rgb}{0.36, 0.54, 0.85}
\definecolor{amaranth}{rgb}{0.9, 0.17, 0.31}
\definecolor{pink}{rgb}{0.87, 0.56, 0.81}
\definecolor{ao}{rgb}{0.0, 0.5, 0.0}
\definecolor{maroon}{rgb}{0.76, 0.13, 0.28}
\definecolor{cardinal}{rgb}{0.77, 0.12, 0.23}
\definecolor{lightcardinal}{rgb}{0.97, 0.42, 0.53}
\definecolor{frenchlila}{rgb}{0.53, 0.38, 0.56}
\definecolor{yellow}{rgb}{1.0, 1.0, 0.87}
\definecolor{lightseagreen}{rgb}{0.45, 0.85, 0.58}
\definecolor{gray}{rgb}{0.9, 0.9, 0.9}
\definecolor{lightblue}{rgb}{0.66, 0.84, 0.96}

\usepackage{color}
\hypersetup{pagebackref=false,
            hyperindex=false,
            citecolor=blue,
            colorlinks=true,
            linkcolor=blue,
            urlcolor=blue}

\newcommand{\eryn}{{\tt Eryn}}
\newcommand{\emcee}{{\tt emcee}}
\newcommand{\param}{\ensuremath{\vec{\theta}} }

\DeclareRobustCommand{\VAN}[3]{#2}
\let\VANthebibliography\thebibliography
\def\thebibliography{\DeclareRobustCommand{\VAN}[3]{##3}\VANthebibliography}

\title[\eryn{}]{\eryn{}\thanks{\href{https://github.com/mikekatz04/Eryn}{https://github.com/mikekatz04/Eryn} } : A multi-purpose sampler for Bayesian inference}

\author[Nikolaos Karnesis et al.]{Nikolaos Karnesis,$^{1}$\thanks{E-mail: \href{mailto:karnesis@auth.gr}{karnesis@auth.gr} }
Michael L. Katz,$^{2}$
Natalia Korsakova,$^{3}$
Jonathan R. Gair,$^{2}$
Nikolaos Stergioulas$^{1}$
\\
$^{1}$ Department of Physics, Aristotle University of Thessaloniki, Thessaloniki 54124, Greece \\
$^{2}$Max-Planck-Institut f\"ur Gravitationsphysik, Albert-Einstein-Institut,  Am M\"uhlenberg 1, 14476 Potsdam-Golm, Germany\\
$^{3}$Astroparticule et Cosmologie, CNRS, Universit\'e Paris Cit\'e, F-75013 Paris, France
}

\pubyear{2023}

\begin{document}
\maketitle

\begin{abstract}
In recent years, methods for Bayesian inference have been widely used in many different problems in physics where detection and characterization are necessary. Data analysis in gravitational-wave astronomy is a prime example of such a case. Bayesian inference has been very successful because this technique provides a representation of the parameters as a posterior probability distribution, with uncertainties informed by the precision of the experimental measurements. During the last couple of decades, many specific advances have been proposed and employed in order to solve a large variety of different problems. In this work, we present a Markov Chain Monte Carlo (MCMC) algorithm that integrates many of those concepts into a single MCMC package. For this purpose, we have built \eryn{}, a user-friendly and multipurpose toolbox for Bayesian inference, which can be utilized for solving parameter estimation and model selection problems, ranging from simple inference questions, to those with large-scale model variation requiring trans-dimensional MCMC methods, like the LISA global fit problem. In this paper, we describe this sampler package and illustrate its capabilities on a variety of use cases. 
\end{abstract}


\section{\label{sec:intro}Introduction}

In physics, and in science in general, one of the most encountered problems is the one of model calibration and comparison. We test our models of the physical world against the measured data, to estimate their parameters and to robustly determine the most suitable model that describes our observations. A crucial first step in this direction, is to efficiently explore the posterior distribution of the parameters given the measured data. Markov Chain Monte Carlo (MCMC) algorithms have proven to be very successful in this regard~\citep{Metropolis1959,Hastings1970,mhhistory, gilks1995markov}, being one of the few methods which can efficiently perform Bayesian inference when the posterior is not analytically tractable and without solving exactly for the marginal likelihood. This is compared to, for example, grid methods, which are often computationally unfeasible. This is especially true in the field of Gravitational-Wave (GW) astronomy, where MCMC methods have been extensively used in order to find physical parameters for signals buried in the data~(see e.g. \citet{bilby,pycbc,LIGOScientific2021djp, Abbott_2020, SHAWHAN2003396, LIGOScientific2019lzm}), as well as to hierarchically infer the properties of the underlying astrophysical populations~(e.g. \citet{hier1, thrane_talbot_2019, Isi2022cii}). MCMC approaches can also compute the {\it marginal likelihood} or {\it evidence} (see section~\ref{sec:mcmc}), by using techniques such as {\it Thermodynamic Integration} (see section~\ref{sec:pt}). In a Bayesian framework, the evidence difference between two models can be used to compute the Bayes Factor, which is used to select between different models that could describe the observations. 

Thermodynamic integration (or other approximations, see section~\ref{sec:evidence} or~\citet{rjmcmc_app1,Gelman1996}), is ideal for cases where the number of competing models is small. However, in situations where the number of potential models becomes too large, the task of iteratively and hierarchically computing the marginal likelihood can become computationally inefficient, or even practically unachievable. Such is the case for future signal-dominated GW observatories, such as the Laser Interferometer Space Antenna (LISA)~\citep{lisa1} or other proposed space-borne  GW observatories~\citep{TianQin, decigo, taiji}. LISA will observe different types of GW sources, the most numerous of them being the Ultra Compact Binaries (UCBs) within the Milky Way~\citep{lisa1, lisa2, rjmcmc_app0, protoglobal, Strub2022upl, Zhang2021, stochlisa, crowder2007}. Those are mostly comprised of a population of Double White Dwarfs (DWD), although a small fraction of neutron star - white dwarf (NS-WD) or double neutron stars (NS-NS) binaries are expected~\citep{breivik2020,Nelemans2001}. In fact, LISA is going to detect GWs from the complete population of $\mathcal{O}(10^7)$ sources simultaneously, but only a small fraction of them are going to be individually resolvable ($\mathcal{O}(10^4)$). The large majority of signals will generate an anisotropic and non-stationary ``confusion'' type of signal, which will dominate the LISA band between $0.05$ and $\sim0.2$~$\mathrm{mHz}$. 

In the above context, computing the marginal likelihood for such a large parameter space and for all possible numbers of events that could be in the population becomes computationally prohibitive. Instead, we can employ {\it dynamical trans-dimensional} MCMC methods~\citep{rjmcmc1}. This family of methods can be quite challenging to tune, but it has proven to yield satisfactory results, even for such demanding problems as the LISA data~\citep{rjmcmc_app0, protoglobal}. There are also implementation challenges, which arise from technical aspects of the algorithm; one example being the {\it dimension matching} requirement when proposing moves between models with different dimensionality. In terms of algorithm efficiency, it is also crucial to choose proposal distributions that allow smooth transitions on a dynamical parameter space, a task which in many cases requires substantial effort. For these technical reasons, all the available software tools of this kind have been specifically developed for the particular problem they intend to solve. 

In this work, we present \eryn{}, a reversible jump MCMC algorithm, capable of {\it efficiently sampling dynamical parameter spaces, while remaining generic and usable by a large community}. We build upon various ideas from statistics, astronomy, etc., in order to develop an efficient statistical toolbox that can be applied to the majority of problems involving detection and characterization of signals. Our primary goal, however, is to utilize \eryn{} as a basic ingredient for a data analysis pipeline to perform the LISA {\it Global Fit}~\citep{rjmcmc_app0, protoglobal}. The Global Fit is a data analysis strategy required to tackle the problem of multiple source detection, separation, and characterization in LISA data. For demonstration purposes in this work, we use \eryn{} to analyze a ``reduced'' scenario of the LISA data in section~\ref{sec:gws}. 

This paper is organized as follows: in section~\ref{sec:mcmc}, we begin with explaining the foundations of the MCMC algorithm, as well as some of the relevant methods that we have adopted for our implementation. In section~\ref{sec:eryn}, we describe how the methods introduced in section~\ref{sec:mcmc} are combined into the actual toolbox implemented in \eryn{}. In section~\ref{sec:examples}, we demonstrate the capabilities of \eryn{} through some toy examples, while in section~\ref{sec:gws} we apply our machinery to more demanding applications in Gravitational-Wave astronomy. In particular, we demonstrate \eryn{} by performing model selection on a simulated population of Ultra Compact Galactic Binaries (UCBs) as measured by the future LISA observatory. Finally, in section~\ref{sec:discussion}, we summarize our work and discuss future applications. We should state again here, that \eryn{} is available as open source software in \url{https://github.com/mikekatz04/Eryn}.

\section{Markov Chain Monte Carlo algorithms}
\label{sec:mcmc}

Nowadays, MCMC methods are considered to be a cornerstone of Bayesian Inference, being very effective in finding solutions to problems encountered across wide-ranging disciplines~\cite[e.g.][]{bayesincosmology, bilby, pycbc, Hogg_2018, astro2017,econometrics}. These include the sampling of the posterior densities of parameters of interest, the numerical marginalisation over nuisance parameters, and providing a framework to compute the marginal posterior distributions (or evidences) that can be used for model selection. The Bayesian framework is based around Bayes' Theorem:
\begin{equation}
	p (\param |y, \mathcal{M}) = \frac{p (y | \param, \mathcal{M}) p(\param| \mathcal{M})}{p(y| \mathcal{M})},
	\label{eq:bayesrule}
\end{equation}
where $y$ is the measured data  and $\mathcal{M}$ our chosen model of analysis. The $p (\param |y)$ term is the posterior distribution of the parameter set $\param$, which is related to the likelihood function of the data $p (y | \param, \mathcal{M})$ and the prior densities of the parameters $p(\param| \mathcal{M})$. The evidence $p(y| \mathcal{M})$ is the marginal posterior over the parameter space $\vec{\theta}\in\Theta$:
\begin{align} 
	\mathcal{Z} \equiv p(\vec{y}| \mathcal{M}) &= \int_\Theta  p(\param , \vec{y}| \mathcal{M})\mathrm{d}\param \nonumber \\
 &= \int_\Theta  p(\vec{y}| \param , \mathcal{M}) p(\param| \mathcal{M}) \mathrm{d}\param.
	\label{eq:evidence}
\end{align}
For parameter estimation purposes, the evidence acts as a normalization constant and can be ignored. However, it is really important if one wants to perform model selection over the measured data. We shall describe in detail how one can numerically approximate the integral of eq.~(\ref{eq:evidence}) in section~\ref{sec:evidence}.

MCMC algorithms work by constructing a Markov Chain sequence, whose elements, $\param(t_i)$, for $i=0, 1, \ldots$, are (asymptotically) independent samples from the target distribution, $f(\param)$. Under fairly general assumptions, the distribution of samples in the chain will converge to the target distribution provided the algorithm satisfies {\it detailed balance}:
\begin{equation}
    f(\param) p(\param \rightarrow \param') = f(\param') p(\param' \rightarrow \param).\label{eq:detbal}
\end{equation}
Here $p(\param \rightarrow \param')$ is the probability that the Markov chain moves from point $\param$ to point $\param'$. The most widely-used MCMC algorithm is Metropolis-Hastings~\citep{Metropolis1959,Hastings1970}, which is explained in algorithm box~\ref{algo:mh}. The first step of the algorithm is to define an initial state, $\param(t_0)$. Then, at each subsequent step $i$, a new state is proposed, by randomly drawing from a given proposal distribution $q(\param' | \param(t_i))$. The newly proposed state is then accepted with a certain probability, given by eq.~(\ref{eq:acceptanceratio}). If the move is accepted we set $\param(t_{i+1})=\param'$, otherwise we set $\param(t_{i+1})=\param(t_i)$. Any reasonable choice of the proposal density will generate a Markov chain with the correct stationary distribution. However, a good choice of $q$ is critical for its efficiency, i.e. achieving the convergence of the MCMC chains within a reasonable computational time. For the special case of a symmetric proposal distribution\footnote{That is, a proposal with the property $p(\param \rightarrow \param') = p(\param' \rightarrow \param)$.}, such as the widely used multivariate Gaussian distribution centered around $\param(t_i)$, the ratio of eq.~(\ref{eq:acceptanceratio}) in algorithm box~\ref{algo:mh} becomes simply the ratio of the target densities at the current $\param(t_{n})$ and proposed $\param'$ points. For high-dimensional problems, the multivariate Gaussian proposal can be tuned during the burn-in period of sampling to improve efficiency~\citep{adaptive_proposal,adaptive_proposal_2,adaptive_proposal_3, Gelman1996, 10.2307/25651249, RevModPhys.94.025001}, or even scaled according to the Cramer-Rao bound, estimated from the Information matrix~\citep{valis2008}. 
\begin{algorithm}
\caption[The Metropolis-Hastings.]{The Metropolis-Hastings algorithm~\citep{Metropolis1959,Hastings1970}. $f$ is the target density to be sampled.}
\begin{algorithmic}[1]
\small
\State Set the initial state of chain: $\param(t_0)\equiv\param$.
\State At each subsequent step $n$, draw a new state $\param'$ from a {\it proposal distribution} $q(\param' | \param(t_{n-1}))$.
\State Take \[\param(t_{n}) = \left\{ \begin{array}{lr}
    \param' \mbox{ with probability } \alpha(\param(t_{n-1}), \param') \\
    \param \mbox{ with probability } 1-\alpha(\param(t_{n-1}), \param') \end{array} \right. ,\]
\item[]  where \begin{equation} \alpha(x,y)= 1 \wedge \left\{ \frac{f(y)q(x|y)}{f(x)q(y|x)} \right\}. \label{eq:acceptanceratio} \end{equation}
\State Go to 2 and repeat until equilibrium is reached, and enough independent samples have been drawn from the target distribution.
\end{algorithmic}
\label{algo:mh}
\end{algorithm}

Although the MH algorithm has been quite successful in tackling inference problems, there are practical implementation issues to overcome. Improving acceptance rate is crucial for convergence, and sometimes improvements in the proposal distribution are  not sufficient to efficiently sample the parameter space. To tackle these issues, various MCMC enhancements have been proposed. A prime example is the Hamiltonian Monte Carlo (HMC) algorithms that utilize local gradients in order to generate proposal points~\citep{hmc1,hmc2}. One variant of HMC is the No-U-Turn sampler which automates part of the required tuning of the HMC~\cite{nuts} sampler. Another alternative to the ``standard'' MH is the Gibbs sampling algorithm, which is particularly useful if the conditional distributions of the parameters of the model are known~\citep{gibbs1, gibbs2,gibbs3}. All of the above developments have been shown to be useful in various disciplines~\citep{bda_gelman, Hogg_2018, brooks, robert, Josephqft, astro2017, kendall2005markov, sivia2006data, sorensen2007likelihood,JOHANNES20101,baio2012bayesian}. Finally, there have recently been numerous proposals that aim to enhance sampling with machine learning techniques. At their core, many of these methods optimize the exploration of the likelihood surface, either by learning it directly (see for example~\citet{likeratio}) or by sampling it in a simpler latent space (for example~\citet{stochastic_nf}).

In this work, we introduce \eryn{}, which is built around the \emcee{} package~\citep{emcee}, enhanced with a variety of sampling mechanisms that allow us to perform inference on dynamical parameter spaces with minimal tuning. We expand on the most important features in the sections below.

\subsection{\label{sec:emcee}Affine-invariant samplers}
An affine transformation is one of the form $\param \rightarrow \vec{\zeta} = A \param + b$, where $A$ and $b$ are a constant matrix and vector respectively. Under an affine transformation a probability density $p (\param |y)$ transforms to
\begin{equation}
	p_{A,\,b} (\vec{\zeta} |y) = p(A^{-1} (\vec\zeta -b) | y)/{\rm det}(A).
	\label{eq:aff1}
\end{equation}
Such transformations can help to transform difficult-to-sample distributions into easier-to-sample ones. A simple example is a multivariate Normal distribution. If the dynamical range of the eigenvalues of the covariance matrix is very large, then sampling can be difficult, but any multivariate Normal distribution can be transformed into a spherical distribution via an affine transformation.

Affine-invariant MCMC is a class of samplers that are designed to have equal sampling efficiency for all distributions that are related by an affine transformation~\citep{emcee, affine1}. The sequence of samples in a Markov chain, $\{X(t)\}$, can be written deterministically as a function of a sequence of random variables, $\xi(t)$, which represent the random draws used to propose new points and evaluate the accept/reject decision. Specifically we can always write
\begin{equation}
    X(t+1) = R(X(t), \xi(t), p)
    \label{eq:MCMCdeterministic}
\end{equation}
where $p$ denotes the target density. An affine-invariant sampler has the property
\begin{equation}
    R(AX(t)+b, \xi(t), p_{A,b}) = AR(X(t), \xi(t), p) + b,
\end{equation}
i.e., the sequence of points visited when sampling an affine-transformed density are the affine transformations of the states visited when sampling the original density. If an affine transformation exists that maps the given target density to one which is more straightforward to sample from, an affine-invariant sampler should sample it as efficiently as it could the simpler distribution, so the convergence  of affine-invariant samplers is less affected by correlations between the parameters~\citep{emcee}. 

In practice, this goal is achieved by following an ensemble of points, called walkers, and basing proposed moves on the distribution of other points in the ensemble. In~\citet{emcee}, the primary update move is the so-called {\it stretch-move} proposal. 
Each walker at state $X_i (t)$ is updated by randomly selecting another walker $j$ and proposing a new value $Y = X_j(t)+ Z[X_i(t) - X_j(t)] $, where $Z$ is a random variable drawn from the distribution~\citep{affine1}
\begin{equation}
    g(z) \propto \left\{ \begin{array}{ll} \frac{1}{\sqrt{z}}& z \in \left[ \frac{1}{a}, a\right]\\0&\mbox{otherwise}\end{array}\right. .
    \label{eq:stretchprop}
\end{equation}
The parameter $a$ can be tuned to improve convergence, but $a=2$ works well in the majority of applications~\citep{emcee}. The proposed point is accepted with probability
\begin{equation}
    \alpha(X_i (t), Y) = 1 \wedge \left\{ z^{d-1} \frac{p(Y)}{p(X_i(t))}\right\}
\end{equation}
where $p$ is the target density and $d$ is the dimension of the parameter space. This acceptance probability is specific to the stretch proposal distribution given by Eq.~(\ref{eq:stretchprop}). For other stretch proposals, the term $z^{d-1}$ must be replaced by $z^{d-2} g(1/z)/g(z)$. Following this scheme, detailed balance is maintained, and it can be proven that affine-invariant samplers converge faster to their target distribution~\citep{emcee}. Below in section~\ref{sec:eryn} we discuss the extension of the stretch-move proposal to Reversible-Jump MCMC methods. The benefits of running MCMC chains in parallel, combined with a proposal distribution that requires almost no tuning, have contributed to an increasing popularity of affine-invariant samplers. In particular, the \emcee{} package~\citep{emcee}, has been used widely in Astrophysics and Cosmology~\citep[e.g][]{virtanen_scipy_2020, 2020MNRAS4981420W, 2017MNRAS46576M, 2018PhRvL121i1102D}.

\subsection{\label{sec:dr}Delayed Rejection}

The Delayed Rejection (DR) scheme of sampling was devised in order to improve two aspects of MCMC algorithms. First, it allows for improvements in the acceptance rate of the proposals, yielding ``healthier'' parameter chains, with better mixing. Secondly, it is more robust against becoming trapped in local maxima of the posterior surface~\citep{dr1,dr2,dr3,dr4}. The strategy, as the name suggests, is, at each iteration, instead of immediately rejecting the newly proposed point based on algorithm~\ref{algo:mh}, we keep proposing new points while maintaining detailed balance by computing both the forward and backward transition probabilities. Suppose we are at a point $\param_0$ and use a proposal $q(\param_1|\param_0)$ to propose a new point $\param_1$. The usual acceptance probability, following the notation of eq.~(\ref{eq:bayesrule}), is 
\begin{equation}
    \alpha_1(\param_0,\param_1)= 1 \wedge \left\{  \frac{p(\param_1 | y ) q(\param_0| \param_1)}{p(\param_0 | y )q(\param_1| \param_0)} \right\}, 
    \label{eq:dr1}
\end{equation}
as per eq.~(\ref{eq:acceptanceratio}). If $\param_1$ is rejected, then instead of going back to step 1 of algorithm~\ref{algo:mh}, we propose a new point, $\param_2$, drawn from a proposal distribution $q(\param_2|\param_1,\param_0)$. This proposal distribution may depend only on $\param_1$, but we write it more generally here to allow for the case that the proposal is adapted based on the sequence of steps that have been rejected. The acceptance probability for $\param_2$, $\alpha_2(\param_0,\param_1,\param_2)$, is computed using 
\begin{equation}
\begin{aligned}
    & \alpha_2 (\param_0,\param_1,\param_2) = \\
    & 1 \wedge \left\{  \frac{p(\param_2 | y ) q(\param_1| \param_2)q(\param_0| \param_1, \param_2) \left[ 1 - \alpha_1\left( \param_2, \param_1 \right)  \right]}{p(\param_0 | y )q(\param_1| \param_0)q(\param_2| \param_1, \param_0)\left[ 1 - \alpha_1\left( \param_0, \param_1 \right)  \right]} \right\}. 
    \label{eq:dr2}
\end{aligned}
\end{equation}
If $\param_2$ is rejected, further steps can be included and each step adds additional proposal and rejection-probability terms to the numerator and denominator of the acceptance probability. For example, the three step acceptance probability, $\alpha_3(\param_0,\param_1,\param_2,\param_3)$ is the minimum of one and
\begin{align}
&\frac{p(\param_3 | y ) q(\param_2| \param_3)q(\param_1| \param_2, \param_3) q(\param_0| \param_1,\param_2, \param_3) }{p(\param_0 | y )q(\param_1| \param_0)q(\param_2| \param_1, \param_0)q(\param_3|\param_2, \param_1, \param_0)} \nonumber \\
&\hspace{0.5cm}\times \frac{\left[ 1 - \alpha_1\left( \param_3, \param_2 \right)  \right] \left[ 1 - \alpha_2\left( \param_3, \param_2,\param_1 \right)  \right]}{\left[ 1 - \alpha_1\left( \param_0, \param_1 \right)  \right] \left[ 1 - \alpha_2\left( \param_0, \param_1, \param_2 \right)  \right]}
\label{eq:dr3}
\end{align}
The proposal $q$ can be different at each step, as long as the relevant proposal density is used in eq.~(\ref{eq:dr2}). For example, in~\citet{dr1} the proposal is built upon a Gaussian mixture model that tries further points in the parameter space with the aim of efficiently exploring multiple modes of the posterior distribution. As the number of steps in the DR scheme becomes arbitrarily large the acceptance probability slowly approaches zero. This algorithm is also limited in practice by high computational requirements, since at every delayed rejection step we need to evaluate a new likelihood and compute the backwards probability (the $\alpha_1 ( \param_2, \param_1 )$ from eq.~(\ref{eq:dr2})). Nevertheless, the DR scheme offers many advantages, and despite the computational cost, it is very useful when the posterior surface exhibits high dimensionality, and when acceleration techniques are available. These, for example, might include the use of Graphical Processing Units (GPUs), and/or heterodyned likelihoods~\citep{2021PhRvD104j4054C}. In our implementation here, we follow closely the one in~\citet{dr1}, for improving the acceptance rate of the {\it between-model step} of the Reversible Jump algorithm (see section~\ref{sec:rj}). As already mentioned, the Reversible Jump MCMC  allows for sampling dynamical parameter spaces. In the special case of nested models, such as the case of searching multiple signals in the LISA data, proposing the `birth' of a signal out of a very wide prior can be very inefficient. A delayed rejection scheme alleviates this problem, by effectively performing a small search around the first set of rejections, increasing the chances of finding a good signal candidate, and thus improving the mixing of the chains.

\subsection{\label{sec:mt}Multiple Try Metropolis}

The Multiple Try Metropolis (MTM)~\citep{mt1,mt2,mt3,mt4} is a sub-class of the implementation of the MH algorithm, which is based on the idea of generating a number of proposals for each individual current state, and then selecting one of them based on their importance weight. In proposing a move from $\param_{t-1}$, a set of $N$ possible new points, $\{y_k\}$, are drawn from a proposal distribution $q(y|\param_{t-1})$ and are assigned weights $w_i = w(y_i | \param_{t-1})$ using a weight function $w(\cdot|\param_{t-1})$. One of these proposed points, $y_J$, is selected with probability given by the normalised weight
\begin{equation}
    \bar{w}_i = \frac{w_i}{\sum_{k=1}^N w_k}.
\end{equation}
To compute the acceptance probability we need to draw $N-1$ points, $\{x_i, i=1,\ldots,N-1\}$, for the reverse move from the proposal $q(x|y_J)$, and assign weights $w(x|y_J)$.
We then set $\param_t = y_J$ with probability
\begin{equation}
\alpha(\param_{t-1},y_J)= 1 \wedge \left\{  \frac{ w(y_J|\param_{t-1}) + \sum_{k=1,k\neq J}^N w(y_k|\param_{t-1}) }{ w(\param_{t-1}|y_J) + \sum_{k=1}^{N-1} w(x_k|y_J) } \right\}.
\label{eq:mtm}
\end{equation}
and set $\param_t=\param_{t-1}$ otherwise~\citep{mt1}. This procedure will satisfy detailed balance if the weight function is chosen such that
\begin{equation}
    p(\param_0|y)q(\param_1|\param_0)w(\param_1|\param_0) = p(\param_1|y)q(\param_0|\param_1)w(\param_0|\param_1).
    \label{eq:MTMdetbal}
\end{equation}
This will be satisfied by a weight function of the form
\begin{equation}
    w(\param_t| \param_{t-1}) = p(\param_t |y) q (\param_{t-1}| \param_{t}) \xi (\param_{t-1}, \param_t),
    \label{eq:w}
\end{equation}
where $\xi (\param_{t-1}, \param_t)$ is any symmetric function, i.e., $\xi (\param_{t-1}, \param_t) = \xi (\param_t, \param_{t-1})$, $\forall \param_t, \param_{t-1} \in {\cal D} \subseteq \mathbb{R} ^d $, with $d$ being the dimensionality of the problem at hand. The detailed balance condition can also be satisfied by a weight function of the form
\begin{equation}
    w(\param_t| \param_{t-1}) = \frac{p(\param_t |y)} {q (\param_{t}| \param_{t-1})}.
\end{equation}
Making this choice and additionally using a proposal function that is independent of the current point, $q (\param_{t}| \param_{t-1}) = q (\param_{t})$ only, we obtain the {\it Independent MTM} algorithm \citep{mt1}. When using the independent MTM algorithm detailed balance is maintained when the same set of points is used for the reverse proposal as for the forward proposal, which saves the evaluation of $N-1$ posterior densities.

The base MTM is currently implemented in {\tt Eryn} with options for the Independent MTM algorithm and symmetric proposals. For a symmetric proposal distribution, $q(\param_{t-1}|\param_t) = q(\param_t | \param_{t-1})$, eq.~(\ref{eq:MTMdetbal}) can be satisfied using the weight function $w(\param_1|\param_0) = p(\param_1|y)$. In this case, we still need to draw separate samples for the reverse step (unlike in the Independent MTM case).


Generating a large number of candidate points yields certain advantages. As expected, the first advantage is the fact that there is usually very good coverage of the parameter space. The second is that the implementation of the MTM usually results in chain states with very low correlation between them. Nevertheless, as for Delayed Rejection, this algorithm requires increased computational resources, since multiple likelihoods/posterior densities have to be evaluated at each iteration of the chain. This cost can be offset in cases where the computations can be parallelized, for example using either CPU or GPU acceleration.

\subsection{Adaptive Parallel Tempering}
\label{sec:pt}

The concept of Parallel Tempering was introduced in order to efficiently sample surfaces with high multi-modality~\citep{pt1, pt2, Vousden2016}. The idea is based on a transformation of the posterior density to a density with a different temperature, $T$, defined by
\begin{equation}
    p_T(\param | y) \propto p(y | \param ) ^{1/T} p(\param).
    \label{eq:pt_basic}
\end{equation}
For $T=1$ this is the target posterior density. In the limit $T\rightarrow \infty$ it is the prior density. Intermediate temperatures ``smooth out'' the posterior by reducing the contrast between areas of high and low likelihood. 

In parallel tempering, a set of Markov chains are constructed in parallel, each one sampling the transformed posterior for a different temperature $T$. These chains periodically exchange information. The idea is that the hottest chains explore the parameter space more widely, and information about areas of high likelihood that they encounter propagate to the colder chains. Information is exchanged by proposing swaps of the states between the different chains. If two chains are sampling from target densities $p_1(\param)$ and $p_2(\param)$ respectively, then the transition probability for chain $1$ in the swap is $p_1(\param_0 \rightarrow \param_1) = p_2(\param_1) \alpha(\param_0,\param_1)$. Detailed balance is thus maintained by accepting the swap with probability
\begin{equation}
    \alpha(\param_0,\param_1) = 1\wedge\left\{ \frac{p_1(\param_1) p_2(\param_0)}{p_1(\param_0) p_2(\param_1)}\right\},
\end{equation}
which for the specific case of swapping between two tempered chains $i$ and $j$ when doing parallel tempering is
\begin{equation}
    \alpha_{i,j} = 1 \wedge \left\{ \left( \frac{p(y | \param_i )}{p(y | \param_j )} \right)^{\beta_j - \beta_i} \right\},
    \label{eq:pt_accrate}
\end{equation}
with $\beta_i = 1/ T_i$ being the inverse temperature, and $\param_i$ the given parameter state for the $i$-th chain.

The temperature ladder $T_i$ should be chosen in order to maximize the information flow between chains of different temperatures, so as to encourage the efficient exploration of the complete parameter space. Typically, this ladder can be static or dynamically adjusted during the sampling procedure. In {\tt Eryn} we have adopted the procedure of~\citet{Vousden2016}, which adapts the temperature ladder based on the swap acceptance rate calculated directly from the chains. Ideally, one should aim for equal acceptance ratio between every pair of neighboring tempered chains, thus tuning their log-temperature-difference $S_i\equiv \log(T_i - T_{i-1})$, according to the swap acceptance rate from eq.~(\ref{eq:pt_accrate}): 
\begin{equation}
    \frac{{\mathrm d} S_i}{{\mathrm d} t} = \kappa (t)\left[ \alpha_{i, i-1}(t) - \alpha_{i+1, i} (t)\right], 
    \label{eq:pt_dt}
\end{equation}
where $\kappa (t)$ tunes the timescale of the evolution of the temperatures. The function $\kappa (t)$ can be chosen depending on the desired behavior of the procedure. In~\citet{Vousden2016} a hyperbolic dependence on the $t$ state is chosen, in order to suppress large dynamic adjustments on long timescales. This setup is the default option in Eryn, but it can be customized. This process is more straightforward for ensemble samplers, where multiple walkers are used, simply because one can get an estimate of the acceptance rate directly from the particular state of the walkers at any given time step $t$. Otherwise, the acceptance rate is computed after iterating for a predefined number of steps, which can be chosen by the user for the given problem at hand.  It can be proven~\citep{Vousden2016}, that the temperature ladder will converge to a particular stable configuration. One should only use this scheme for the initial burn-in stage of sampling, and then continue with a stationary ladder for the rest of the analysis. 

\subsection{\label{sec:evidence}Marginal posterior calculation for model selection}

One of the most frequently encountered problems in physics, and in science in general, is that of model or variable selection, i.e., identifying the model best supported by the observed data. Working in a Bayesian framework, comparison between different hypotheses may be done by computing their evidences or marginal posteriors~\citep{bda_gelman} and evaluating the Bayes Factor:
\begin{equation}
B_{12} = \frac{p(\vec{y}| \mathcal{M}_1)p(\mathcal{M}_1)}{p(\vec{y}| \mathcal{M}_2)p(\mathcal{M}_2)},
\label{eq:bf}
\end{equation}
where the term $p(\mathcal{M}_i)$, is the prior probability assigned to the model ${\cal M}_i$.

The marginal posterior density, or evidence, is given by the integral of eq.~(\ref{eq:evidence}) and is in general quite challenging to compute. For some high signal-to-noise ratio (SNR) cases it can be approximated if the covariance matrix $\Sigma$ of the parameters for all candidate models $\mathcal{M}$ are known. This approach is called the {\it Laplace approximation}~\citep{bayesfactors, bda_gelman}.  However, this is only an approximation, and it sometimes fails for models with weak support from the data~\citep{rjmcmc_app1} (in particular when the posterior cannot be approximated by a multivariate Gaussian at $\param_\mathrm{MAP}$). 

When using parallel tempering~\ref{sec:pt}, it is possible to compute the evidence by a procedure known as {\it thermodynamic integration}~\citep{thermo1}. We define a continuous distribution of evidences based on the target distribution for a chain with inverse temperature $\beta=1/T$ via
\begin{equation}
    \mathcal{Z}_{i,\beta} = \int p(y|\param,\mathcal{M}_i)^\beta p(\param) \, {\rm d}\param.
\end{equation}
For $\beta=0$ the chain is sampling the prior and therefore $\log \mathcal{Z}_{i,0}=0$. For $\beta=1$ we are sampling the target distribution and $\log\mathcal{Z}_{i,1}=\log\mathcal{Z}_i$. Additionally we have
\begin{align}
    \frac{{\rm d}\log\mathcal{Z}_\beta}{{\rm d}\beta} &= \int \log [p(y|\param,\mathcal{M}_i)] \, p(y|\param,\mathcal{M}_i)^\beta p(\param) {\rm d}\param \nonumber \\
    &\equiv \mathbb{E}_\beta [\log p(y|\param,\mathcal{M}_i)].
\end{align}
From this we deduce
\begin{equation}
    \log \mathcal{Z}_{i} =\int_{0}^{1} \mathbb{E}_{\beta}\left[\log p\left(y \vert \param, \mathcal{M}_{i} \right) \right] d\beta.
    \label{eq:TI_diffsum}
\end{equation}
The expectation value is over the distribution being sampled by the chain at temperature $\beta$ and so can be computed by averaging over the posterior samples~\citep{thermo1, thermo0, Vousden2016}. The integral can then be evaluated using standard methods, for example the trapezium rule, using the full ladder of temperatures. This approach generates reliable evidences, with accuracy limited only by the number of temperatures being sampled, and the efficiency of the sampling of the parameter space $\Theta$ by the chains. Since its first introduction, there have been many applications of this approach, and in particular, there is extensive usage in GW astronomy~\citep{Vousden2016, katz2022, thermo2, thermo3, thermo4}.

The thermodynamic integral in Eq.~(\ref{eq:TI_diffsum}) can be thought of as computing the evidence as a sum of differences between the evidences at different temperatures. An alternative approach, called the {\it stepping-stone algorithm}~\citep{10.1093/sysbio/syq085}, writes the evidence as a product of the ratios of evidences at different temperatures
\begin{equation}
\mathcal{Z}_{i} \equiv \frac{\mathcal{Z}_{i,1}}{\mathcal{Z}_{i,0}} = \prod_{k=1}^{N_T-1} \frac{\mathcal{Z}_{i,\beta_{k+1}}}{\mathcal{Z}_{i,\beta_k}}
\end{equation}
where $\beta_k$ denotes the inverse temperature of chain $k$, $N_T$ is the number of different temperatures being sampled, and we assume $\beta_1 = 0$ and $\beta_{N_T}=1$. Each evidence ratio can be written as a posterior integral
\begin{align}
\frac{\mathcal{Z}_{i,\beta_{k+1}}}{\mathcal{Z}_{i,\beta_k}} &= \int p(y | \param, \mathcal{M}_i)^{\beta_{k+1} - \beta_{k}} \frac{p(y | \param, \mathcal{M}_i)^{\beta_k} p(\param)}{\mathcal{Z}_{\beta_k}} \, {\rm d}\param \nonumber \\
&= \mathbb{E}_{\beta}\left[p\left(y \vert \param, \mathcal{M}_{i} \right)^{\beta_{k+1} - \beta_{k}} \right] \nonumber \\
&\approx \frac{1}{n} \sum_{i=1}^n p\left(y \vert \param^i_k, \mathcal{M}_{i} \right)^{\beta_{k+1} - \beta_{k}},
\end{align}
where $n$ is the number of posterior samples in each chain, and $\param^i_k$ denotes the $i$'th sample at temperature $\beta_k$. This leads to the final expression
\begin{equation}
    \log \mathcal{Z} = \sum_{k=1}^{N_T-1}\log \sum_{i=1}^n p (y | \param^i_k, \mathcal{M})^{\beta_{k+1} - \beta_{k}} - (N_T - 1) \log n .
\end{equation}
In challenging situations, e.g., where the number of tempered chains is relatively small, the stepping-stone algorithm has been shown to produce more accurate estimates of the marginal likelihood than methods that use thermodynamic integration~\citep{10.1093/sysbio/syq085}. This was also demonstrated with practical examples drawn from Gravitational Wave Astronomy in~\cite{thermo2}. 

\subsection{\label{sec:rj}Reversible Jump}

Another approach to the model selection problem is to follow a Reversible Jump (RJ) MCMC strategy, which can dynamically estimate the most probable hypotheses given the data~\citep{rjmcmc1}. The RJ-MCMC is a generalization of the MH algorithm that allows trans-dimensional proposals. Thus, the model order is considered a free parameter which is fitted together with the parameters of the individual models. The most widespread variation of the algorithm uses a two-stage procedure. The first stage or {\it in-model step}, uses the standard MH algorithm to update all the parameters $\param_k$ for the given model $k$. The second stage or {\it between-model step} proposes to update the model state $k$ to a new model state $l$. Parameters $\param_l$ for the new model are also proposed. The newly proposed state $l$, is accepted with a probability defined by~\citep{Godsill} 

\begin{equation}
	\alpha' = 1 \wedge
    \left\{\frac{p(\param_l | l, y ) g'(u_l)  }{p(\param_k | k, y ) g(u_k) } |{\bf J}| \,\right\},
	\label{eq:rjmcmcaccratio}
\end{equation}
where
\begin{equation}
	p(\param_k | k, y ) = p (y | \param_k, k) p(\param_k, k) p(k)
	\label{eq:completepost}
\end{equation}
with $p (y | \param_k, k)$ the likelihood for model $k$, $p (\param_k| k)$ the prior on the parameters $\param_k$ in model $k$, and $p(k)$ the prior for the model state $k$. The term $g'(u_l)|{\bf J}|/g(u_k)$ arises because of the need for {\it dimension matching} between the different model states. In general, we can define a move between model states in terms of a deterministic invertible mapping, $\param_k=q(\param_l,u_l)$ with inverse $\param_l=q'(\param_k,u_k)$, that is a function of the parameters and two sets of random variables, $u_k$ and $u_l$, drawn from distributions $g(u_k)$ and $g'(u_l)$~\citep{Godsill, rjmcmc1}. The term $|{\bf J}|$ is the Jacobian defined by this invertible mapping 
\begin{equation}
    |{\bf J}| = \left|\frac{\partial (\param_l,u_l)}{\partial (\param_k,u_k)}\right|
    \label{eq:jacobian}
\end{equation}
and the term $g'(u_l)/g(u_k)$ plays the role of the proposal ratio in the standard MH acceptance probability. Dimension mapping means that ${\rm dim}(\param_l)+{\rm dim}(u_l)={\rm dim}(\param_k)+{\rm dim}(u_k)$.


Using RJ-MCMC introduces additional computational cost at each MCMC iteration, as well as technical challenges in implementation. Luckily, implementation can be easier when sampling {\it nested models}. This refers to problems where more complicated models contain their simpler counterparts. Examples of such cases are fitting polynomial models, which differ only in the highest order to be determined, or detection problems where multiple similar signals are potentially present the data. In such cases, the between-model step can always be formulated such that the Jacobian of eq.~(\ref{eq:jacobian}) becomes unity, and eq.~(\ref{eq:rjmcmcaccratio}) simplifies to the ratio of posteriors accounting for any differences in prior and proposal volumes~\citep{Dellaportas2002, Lopes2004, rjmcmc_app0}.

After running RJ-MCMC, the Bayes Factor can be approximated by the ratio of the number of iterations spent within each model:
\begin{equation}
	B_{\rm 12} = \frac{\rm \# \:of\: iterations\: in\: model\: {\mathcal M}_1}{\rm \# \: of\: iterations\: in\: model\: {\mathcal M}_2}.
\label{eq:B}
\end{equation}
This algorithm has proven to be robust for evaluating high-dimensional competing models, and has been quite successful in tackling data analysis problems in GW astronomy~\citep{rjmcmc_app0, rjmcmc_app1, rjmcmc_app2} as well as areas spanning physics and signal processing~\citep[e.g.,][]{NIPS1997_0ed94223, 1512059,Yu_2021}. However, designing an efficient RJ-MCMC algorithm can be quite challenging. The first challenge is to choose suitable proposal distributions, which can greatly affect the convergence of the algorithm. In situations where the models are nested, it is both tempting and convenient to take the proposal to be the same as the prior distribution of the parameters. As an illustrative example, we refer to section~\ref{sec:gauss_free}, which describes a toy problem of searching for Gaussian pulses in noisy data. There, the parameters of the individual pulses are the amplitude and location of the pulse described by their $(x,\, y)$ coordinates. In order to search for those signals, the prior on their location must be wide enough to include the complete data set (see figure~\ref{fig:2d_gauss_data}). A birth proposal based on the prior would inevitably be quite inefficient, simply because the chance of proposing a good source candidate is small, especially if the proposal distribution is flat across $(x,\, y)$. We treat the above problem as a motivation to adopt efficient proposals with minimal tuning in \eryn{}, which we further discuss in section~\ref{sec:eryn}. The second major challenge, which of course depends on the given problem at hand, arises from the samplers' capability to explore a multi-modal dynamical parameter space. We discuss our strategy to overcome that challenge in section~\ref{sec:eryn}.  

\section{\eryn{}: Gathering all the pieces together}
\label{sec:eryn}

All the different algorithms described in previous sections can be extremely useful in tackling different kinds of problems that require sampling. In Gravitational Wave Astronomy, we encounter such problems far too often, where dynamical parameter spaces require vast computational resources in order to be explored efficiently. Motivated by those problems, we have implemented a new toolbox that combines all these techniques to enhance the capabilities of an MCMC sampler. We have named this package \eryn{}~\footnote{Eryn Vorn (Sindarin for Blackwood) was a wooded cape in Eriador, and a region of dark pine trees. Located in western Minhiriath, Eryn Vorn (likely named so by the N\'{u}men\'{o}reans ) was originally part of the vast ancient treescape that covered most of north-western Middle-earth}, borrowing the name from the Tolkien mythos~(\citeauthor{eryn}). The analogy has its basis in the idea of a forest: within a forest you have trees which correspond to different walkers, a.k.a. {\it Ents}, in an ensemble MCMC sampler. On each tree there are branches that represent the various types of models used to fit the data. For example, in the case of GW global fitting for LISA, we can imagine using the Galactic binaries as one branch and massive black hole binaries as another branch. Each branch has leaves, which represent the individual instances of each model. In the LISA example, leaves would represent the individual Galactic binaries or massive black hole binaries. And finally, to zoom out, when adding tempering capabilities, we can think of groups of walkers in each temperature taking the form of many forests (of walkers) located within different temperature climates. 

We adopt the architecture of ensemble samplers, and in particular the one of \emcee{}~\citep{emcee}. Having multiple walkers running in parallel is ideal for efficiently sampling the parameter spaces using techniques such as parallel tempering, as described in~\ref{sec:pt} (also see section~\ref{sec:emcee}). In this setting, we evolve $n_w$ walkers per temperature $T_i$, where each walker follows a Reversible Jump MCMC (see section~\ref{sec:rj}), mapping a parameter space of altering dimensionality. In practice, walkers in higher temperatures sample the dynamic parameter space with fewer model components as the penalty from higher prior volume is not compensated by the smoother annealed likelihood. In other words higher temperatures have a sharper Occam's razor: the data can be explained with models that are simpler, or lower-dimensional. The highest temperature chain samples the prior on the model space (provided that $T_\mathrm{max}=\infty$). More details will be given in section~\ref{sec:examples}.

As already mentioned in section~\ref{sec:rj}, Reversible Jump algorithms are extremely challenging to tune, even for simpler classes of problems. One of the major challenges is the low acceptance rate for the between-model proposal, i.e., when we propose a new state where the parameter dimensionality differs. In cases of signal search and detection (which is a nested model situation), it is convenient to set the proposal corresponding to a ``birth'' move to be the same as the prior distribution. In order to accommodate all possibilities for the signals present, the prior densities are usually quite wide, and thus accepting a new higher dimensional state becomes quite improbable. For that reason, within \eryn{}, we have implemented a Delayed Rejection scheme with the aim of improving this acceptance ratio. When proposing $\param_l$ for a higher-dimensional model $l$, we do not reject immediately, but rather make new delayed rejection proposals around the first rejected point $\param_l$, using the given {\it in-model step} update proposal. This, in principle, allows the sampler to explore around $\param_l$ before rejecting the new state~\citep{dr1}, which in turn improves the {\emph between-model step} acceptance rate and produces healthier MCMC chains.

The Delayed Rejection scheme, as described in section~\ref{sec:dr}, requires a serial computation of the delayed rejection acceptance ratio for walkers where the newly proposed state has been rejected. This scheme of calculating costly likelihoods sequentially in a loop during the between-model step, can lead to a computational bottleneck of the MCMC process. This is especially true for the LISA Global Fit problem, where multiple binary waveform signals are present in the data-stream. Then, the computational time for each RJ-MCMC iteration is significantly increased, since the progress will be halted until all walkers have gone through their respective Delayed Rejection process, which requires evaluation of new waveforms at each step. For the reasons summarized here, we have not used the Delayed-Rejection scheme for our analysis in section~\ref{sec:gws}, and have resorted to the Multiple Try scheme. However, the Delayed-Rejection scheme, as explained in section~\ref{sec:dr}, has been implemented in the \eryn{} package.

The Multiply Try scheme was essentially implemented in order to facilitate use of a parallelized likelihood framework. Parallelization is naturally compatible with Multiple Try MCMC as multiple proposals are made for each individual walker, which allows for the parallelized evaluation of proposal distributions, likelihood functions, and acceptance probabilities. Under these parallelized settings, one proposal can act as many when compared to the usual serial evaluation of proposals, allowing for better chain mixing in situations where proposals are infrequently accepted. That being said, it is still important to choose a good proposal distribution, for both the in-model and between-model RJ-MCMC steps, which we discuss further below.

\subsection{\label{sec:proposals}Choosing efficient Proposal distributions}

In the previous sections, we briefly discussed some of the challenges in choosing efficient proposal distributions for both the in-model and between-model steps of the RJ algorithm. For the in-model case, the challenge arises from the fact that it is sometimes impractical, or even unfeasible, to define a well-tuned proposal for each of the possible models that could represent the data. Using again the example of LISA data, one would need to tediously design an effective proposal distribution for the thousands of overlapping binary signals in the data. On the other hand, for the case of the between-model step, choosing proposals from the prior distribution, especially if it is  highly uninformative, can be very inefficient for Reversible Jump sampling. For \eryn{}, in order to tackle those issues, we have implemented the {\it Group Proposals} explained below for addressing the within-model proposals in RJ, as well as a scheme to design an efficient proposal for birth moves during the between-model step, which is based on normalizing flows. 

\subsubsection{\label{sec:groupprop}Group Proposals}

In section~\ref{sec:emcee}, the stretch-move proposal was introduced and discussed. One of the obvious advantages of such a scheme of proposing new MCMC samples is that it requires minimal tuning~\citep{emcee}. However, it does not extend well in its simplest form to the generalized Reversible Jump MCMC. The stretch proposal is based on the idea that the ensemble of points ($X_j$) is sitting on the same posterior mode as the current point ($X_i$). In a nested model situation where both the model count and the individual model parameters change, each point may lie on a different posterior mode representing a different set of leaves (sources) in the data. This can be alleviated by applying the stretch move to individual leaves within each branch of each walker, but there is still an issue of identifying leaves in different walkers that lie on the same posterior mode. The stretch proposal will technically still work when mixing leaves in different posterior modes, but the acceptance fraction will be negatively affected. However, within the stretch proposal formalism, the choice of $X_j$ is customizable. The key to maintaining detailed balance is that $X_j$ cannot depend on $X_i$, and $X_j$ cannot be updated in the same iterative step as $X_i$ \citep{affine1}. 

We leverage this property to design a new type of stretch move that can handle Reversible Jump setups while maintaining a small number of tuning parameters. We call this proposal the ``group'' proposal. The mathematics that govern the group proposal are equivalent to that of the original stretch proposal. The key difference is that the group proposal chooses $X_j$ (see Section~\ref{sec:emcee}) from a stationary group that is fixed for many proposed updates. This is in contrast to the original stretch proposal that always uses the current set of points in the ensemble sampler to draw $X_j$.  
The stationary group is updated after a large number of sampler iterations and we make sure that detailed balance is maintained during the update. We update every leaf within every branch of every walker at each iteration and repeat many iterations between updates of the stationary group.

The appropriate stationary group varies from problem to problem. The goal is to set a group that resembles as best as possible the posterior modes of the current leaves and then draw from it strategically so that the drawn point is likely (but not guaranteed) to lie on the same mode as the leaf that is currently being updated, $X_i$. In the example of the LISA Galactic binaries analysis, we set our stationary group to the full set of leaves (binaries) across all walkers at a specific temperature of the sampler at the end of a given iteration. Then, at proposal time, we efficiently locate the $\sim n_w$ points in the stationary group that are closest to $X_i$ from  based on their initial frequency parameter. We then draw $X_j$ from this group. The hope is that some percentage of the $n_w$ drawn points will lie on the posterior mode on which $X_i$ sits. The exact percentage will vary depending on how close the posterior modes are to each other and how many model instances exist in the sampler that include this specific mode. 
For low SNR binaries, for example, a source may exist in some walkers and not others, making it harder to map its posterior mode with the current group of stationary points. 

The performance of group proposals is highly situation- and/or model-dependent. With individual source posterior modes that are well separated and easy to define in terms of separation, the performance will approach the performance of the base stretch proposal in non-Reversible Jump MCMC because the stationary group will well represent the specific posterior mode on which $X_i$ is located. As the parameter space becomes more crowded and/or separation (distance) metrics become harder to define, the performance of group proposals will worsen.

\subsubsection{\label{sec:groupprop}Learning from the data}

The second improvement concerns the between-model step of the RJ-MCMC. As mentioned earlier, for the case of nested models, it is often convenient to draw ``birth'' candidates directly from the prior distribution of parameters of the given model. This practice can be quite ineffective in terms of acceptance rate. As an example we can again use the LISA data-set case. The UCBs are distributed within the Galactic disk, congregated mostly around its Center~\citep{LISAAstroWorkingGroup}, therefore, adopting a proposal based on an uninformative uniform prior across the sky, would waste computational resources exploring a part of the parameter space with low probability mass. A proposed solution is to use an informative prior derived from the spatial distribution of binaries in the Galaxy~\citep{rjmcmc_app0}. In our work here, we have chosen an alternative data-driven route, based on the actual residual data after a burn-in period of the RJ-MCMC, which we describe below.
\begin{figure*}	
	\centering
	\begin{subfigure}[t]{2.2in}
		\centering
		\includegraphics[width=1.0\linewidth]{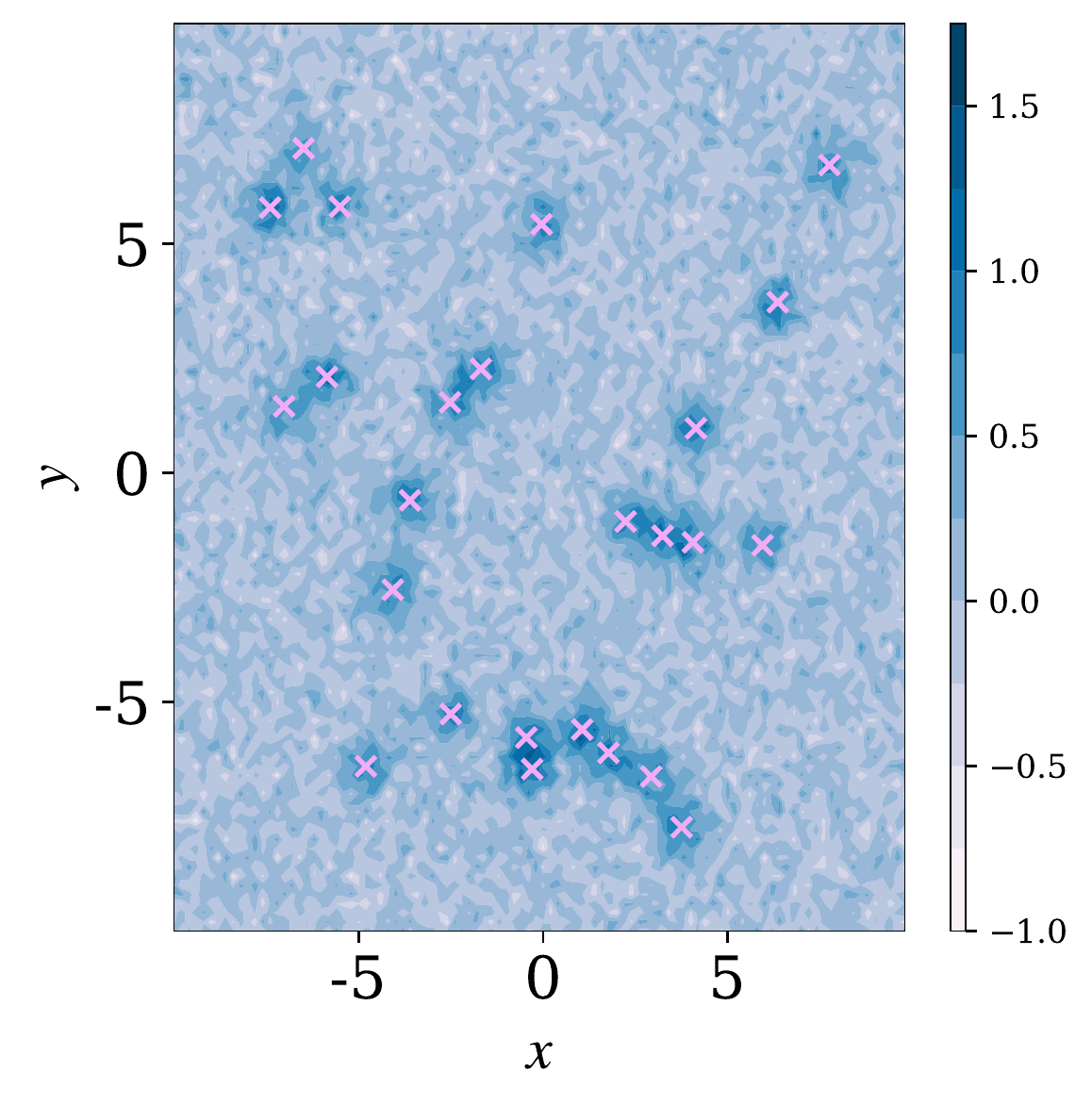}
		\caption{}\label{fig:2d_gauss_data}		
	\end{subfigure}
	\begin{subfigure}[t]{2.2in}
		\centering
		\includegraphics[width=.93\linewidth]{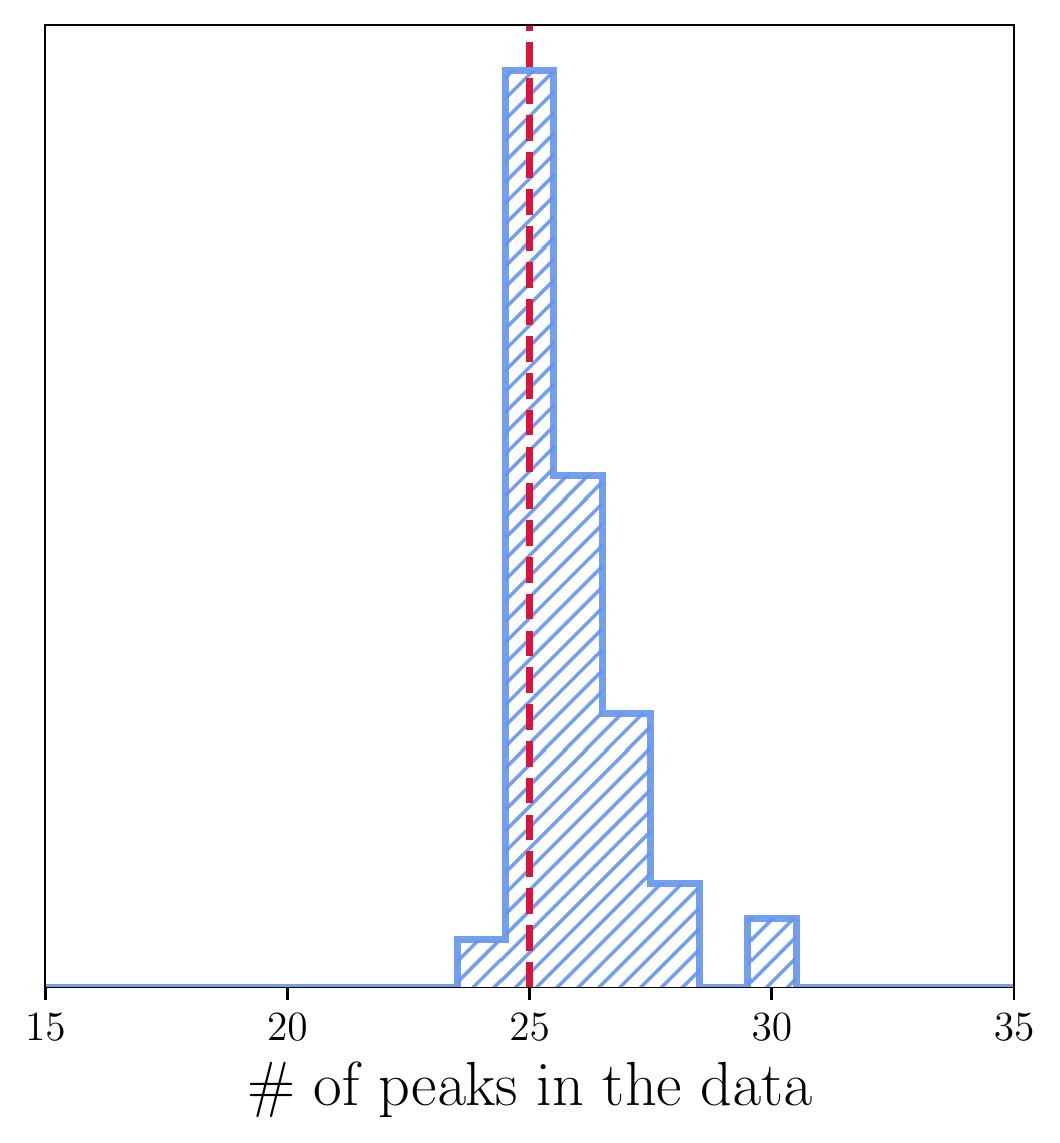}
		\caption{}\label{fig:2d_gauss_hist}
	\end{subfigure}
	\begin{subfigure}[t]{2.2in}
		\centering
		\includegraphics[width=1\linewidth]{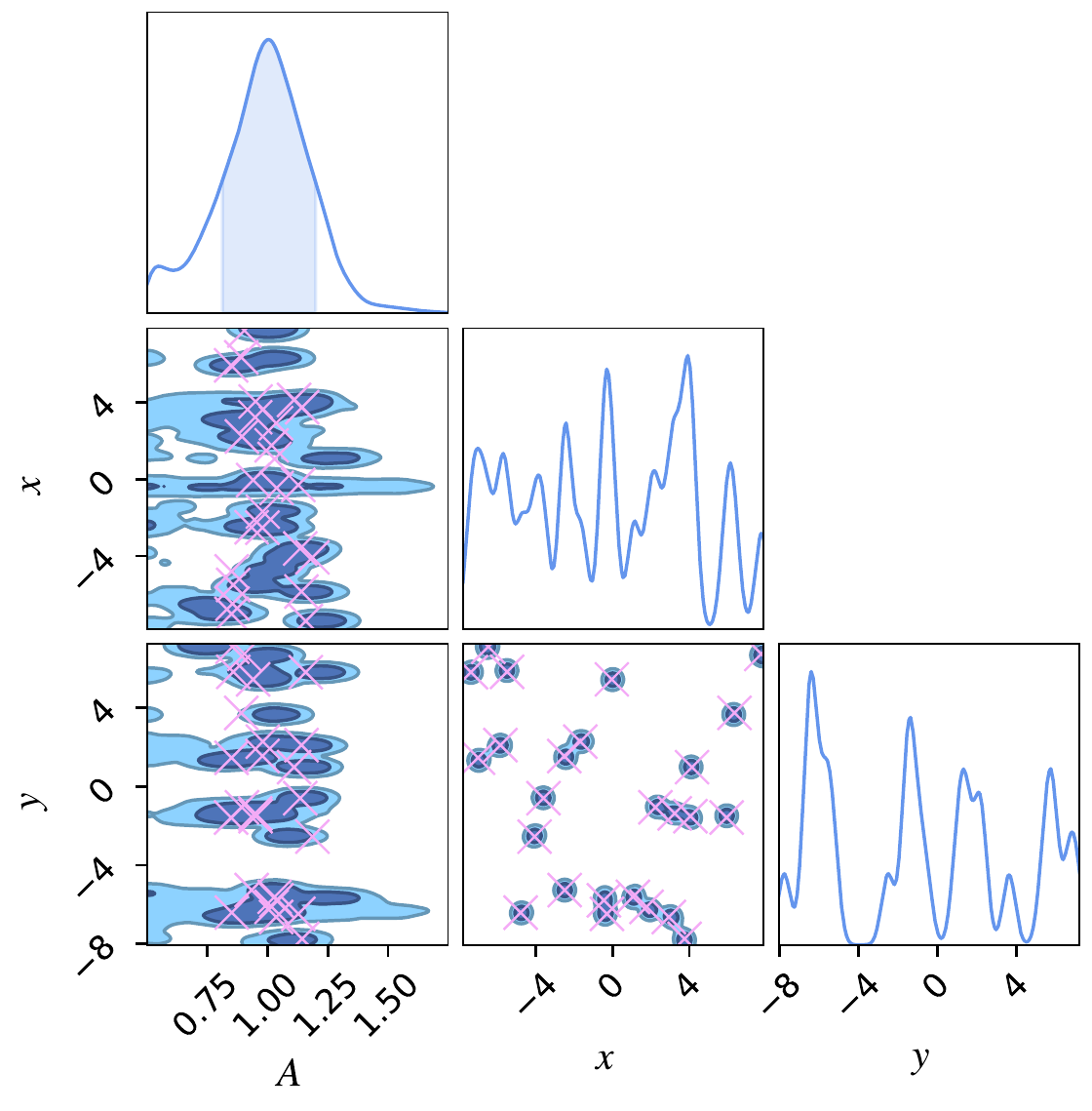}
		\caption{}\label{fig:2d_gauss_posteriors}
	\end{subfigure}
	\caption{Searching for 2D Gaussian pulses in the presence of Gaussian noise. Panel (a): the simulated data, which consists of injections of $25$ pulses in Gaussian noise with $\sigma_n=0.2$. Panel (b): the distribution of the model order, obtained by exploring the dynamical parameter space with \eryn{}. The true value is marked with a dashed red line. For this toy investigation, the correct number of simulated components is recovered. Panel (c): the 2D posterior densities for the parameters of the $k$ Gaussian peaks (see text for details).  }\label{fig:2d_gauss}
\end{figure*}
After a sufficient number of RJ-MCMC iterations, we can extract a subset of sources from nested models which are constantly present in almost all walkers of our cold chain. In other words, we can find and subtract the brightest sources from the data, and then allow for another burn-in period on the resulting residuals. This allows the sampler to explore the remaining parameter space more easily, thus providing a good initial estimate for the weak signals possibly buried in the noisy residual data. We can then use those samples to construct a proposal density which will help us search for good candidates for those weak signals, without excluding the rest of the parameter space. This can be accomplished by  fitting the distribution to the residual data described above. The most efficient way to fit to the generic distribution, is to use an invertible transformation, such as a normalizing flow \citep[for example][]{realNVP, NSF}. The method works in the following way: we sample from the base distribution (which is usually chosen to be Normal $\mathcal{N}(\mathbf{z};0,1)$) and transform samples to the desired distribution $p(\boldsymbol{\theta})$ by applying the change of variable equation 
\begin{equation}
    p(\boldsymbol{\theta}) = N(f^{-1}(\boldsymbol{\theta}))
    \vert J_{f^{-1}} (\boldsymbol{\theta}) \vert .
\end{equation}
Here function $f(\mathbf{z})$ is a bijection which we fit by optimizing the Kullback-Leibler divergence, $D_{\text{KL}}[p(f(\boldsymbol{\theta}))||\mathcal{N}(\mathbf{z};0,1)] $, between a normal distribution and the inverse transform of the distribution that we want to estimate. After the fit has converged we can draw samples from the Normal distribution and transform these to samples from the distribution fitted to all residuals and use it as a proposal. We will cover this method in more detail in a separate paper which is being prepared.

\subsubsection{Convergence Diagnostics\label{sec:convergence}}

Many standard approaches for assessing convergence of MCMC chains can be applied to the RJMCMC chains generated by \eryn{}. Trace plots, both for individual parameters within models, and the indices labelling different models, will show that the chains are mixing well and that the initial burn-in phase has finished. Repeating runs with different numbers of walkers, different starting positions, different numbers of parallel chains or different choices of temperatures and comparing results can be used to assess convergence. Similarly, posteriors produced from randomly selected subsets of chain points can be compared (using  for example statistical tools such as the Jensen-Shannon divergence~\citep{MENENDEZ1997307}). 

Additionally, \eryn{} computes and outputs the Potential Scale Reduction Factor (PSRF) for each tempered chain, which provides a quantitative diagnostic of the convergence of the result~\citep{psrf1,psrf2}. For a single parameter, the PSRF index is given by
\begin{equation}
    \widehat{R} = \sqrt{ \frac{(d+3)\widehat{V}}{(d+3) W } } ,
    \label{eq:psrf}
\end{equation}
where $W$ the mean of the empirical variance within each chain, and $\widehat{V}$ is the estimated variance of the all chains, assuming that the target distribution is Gaussian. The degrees of freedom are estimated  by the method of moments as $d \approx 2 \widehat{V} / \widehat{\mathrm{var}}(\widehat{V})$~\citep{psrf2}.

The PSRF index takes values $\widehat{R} \geq 1$, with values close to unity indicating converged MCMC chains. In the end, the quantity that really matters is the PSRF value for the cold chain, but converged chains across for all temperatures is a good indicators for healthy mixing and efficient temperature swaps.

\subsection{\label{sec:implementation} Implementation}

In this section, we discuss the main implementation details of \eryn{}. We refer the interested reader, or user, to the \eryn{} documentation for more exhaustive information and examples~(\citeauthor{eryn_documentation}).

The goal of \eryn{} is to produce a sampler that can handle all (or most) cases of MCMC sampling ranging from basic, non-tempered, single-model type, single-model instance posterior estimation to the full reversible jump MCMC with tempering, multiple model types, and adjustable model counts, as well as everywhere in between. In the basic case, \eryn{} aims to be a close replica of \emcee{} trying to maintain as much simplicity as possible. At the complicated end of the spectrum, \eryn{} attempts to provide a common interface and underlying infrastructure for the variety of problems that may arise, allowing the user to maintain usage of the majority of the code from project to project, focusing on changing only the specific parts of the code that are difficult to implement or require special treatment for each specific problem. Since \eryn{} is effectively an enhanced version of the \emcee{} package, the overall  structure of \emcee{} is strongly maintained. Like in \emcee{}, ``State'' objects move coordinate and likelihood information around the ensemble sampler, storing information in a similar back end object either in memory or {\tt HDF5} files. Additionally, the interface used for adding proposals has remained.

The various enhancements discussed in this work, including tempering, reversible jump moves, multiply try MCMC, etc., are all implemented within the \emcee{}-like structure. This involved two main changes. First, the State objects have been scaled to hold information necessary for reversible jump MCMC: temperature information, prior information, and efficient and concise containers for multiple types of models with an adjustable number of individual model instances. Second, the reversible-jump proposal has been added as a proposal base, similar to the use of the `MH' or `RedBlue' moves within \emcee{}. Beyond these main enhancements, there are also a variety of smaller, but useful, additions to \eryn{} that help the user build a variety of analysis pipelines. These include stopping or convergence functions, functions to periodically update the sampler setup while running, objects to carry special information through the sampler, and aids for coordinate transformation.

\subsection{Toy Examples}
\label{sec:examples}

In this section, we present a series of working examples for \eryn{}. We begin with simple problems, such as searching for simple signals in noisy data, with the aim of demonstrating the performance of this toolbox in a dynamical parameter space. The impact of the different enhancements discussed in section~\ref{sec:eryn} will be assessed and discussed. Finally, in section~\ref{sec:gws}, we will apply this machinery to more realistic problems in Gravitational-Wave astronomy.

\subsubsection{\label{sec:gauss_free} Searching for pulse signals in Gaussian noise}

In this first example, we explore the capabilities of \eryn{} in a simplified application, commonly encountered in physical sciences. We perform a search for Gaussian pulses in a simulated 2D data-set, in the presence of Gaussian noise with variance $\sigma_n = 0.2$. We generate $25$ pulses randomly distributed on the $x-y$ plane with all pulses contained within $x,y \in [-10,10]$ (see figure~\ref{fig:2d_gauss_data}), and amplitude uniformly drawn from $\mathcal{U}[0.7,\, 1.5]$. The amplitude $A_k$ of each pulse, labelled by $k$, is considered a free parameter to be estimated, in addition to the Cartesian coordinates of their centres. The pulses' width was kept fixed to $\sigma_p\times  \delta_{ij}$, with $\sigma_p=0.2$, for the sake of simplicity. Thus, we are required to estimate $N_p$, the total number of pulses in the data, and also estimate the parameters for each individual signal $k$: $\param_k = \{ A_k,\, x_k,\, y_k \}$. The noise variance $\sigma_n$ is estimated as part of the fit.
\begin{figure*}
 	\includegraphics[width=.95\linewidth]{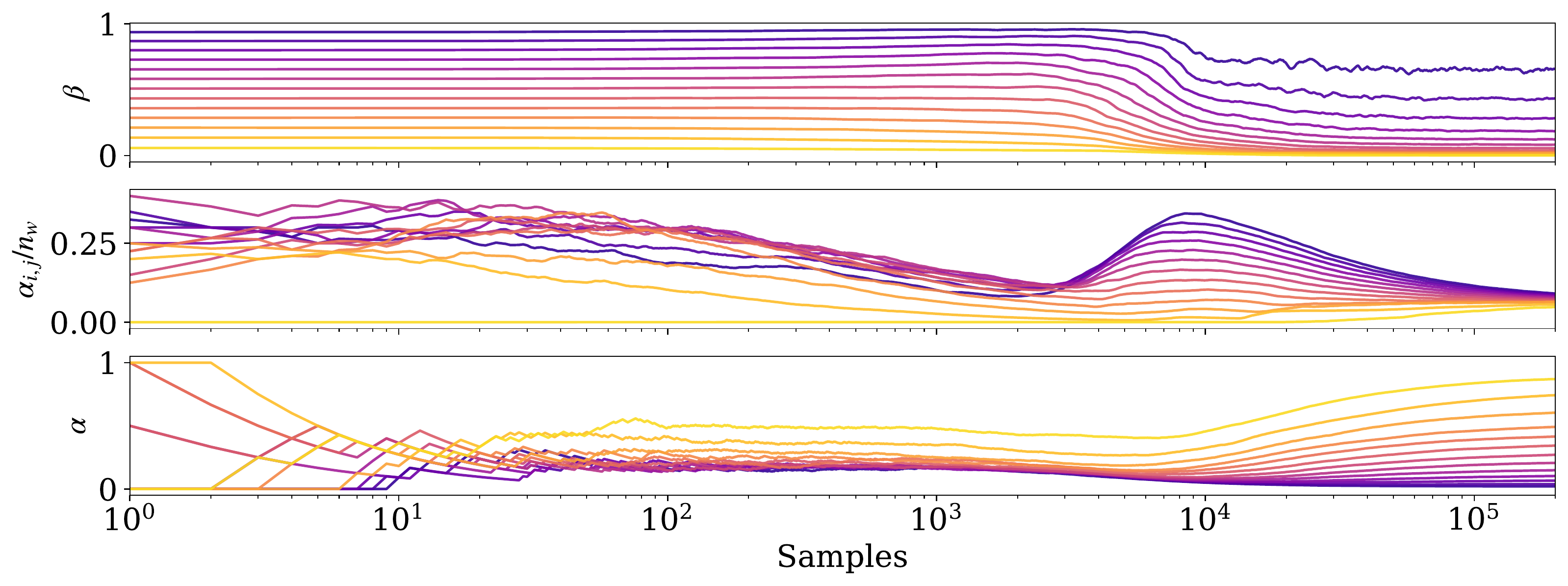}
 	\caption{ {\it Top panel}: The evolution of the temperature chains running in parallel for the toy problem of searching for 2D Gaussian pulses in Gaussian noise. The different colors indicate the the initial temperature chain index. Following the parallel tempering recipe of~\citep{Vousden2016}, the temperature ladder is tuned according to eq.~(\ref{eq:pt_dt}), and the chains start to converge after $\sim10^4$ iterations. {\it Middle panel}: The evolution of the swap acceptance rate $\alpha_{i,j}$ described in eq.~(\ref{eq:pt_accrate}), per number of walkers $n_w$. For this run we have used $n_w=10$ walkers. After $10^5$ iterations, the swap acceptance rate converges to a single (different) value for every temperature chain. {\it Bottom panel}: The ``in-model'' acceptance rate per temperature chain, given by eq.~(\ref{eq:acceptanceratio}). }
\label{fig:gauss_temp_evolution}
\end{figure*}
The analysis of this problem is performed using the adaptive parallel tempering scheme of section~\ref{sec:pt} and the Reversible Jump MCMC proposals (section~\ref{sec:rj}). The in-model proposals for each model component are Gaussians, with a diagonal covariance matrix  $\Sigma=10^{-4} \delta_{ij}$. This proposal is not tuned during sampling. The priors for the parameters are quite wide, covering the entire range of the data, while the prior on the number of pulses $k$ is set to $k\sim\mathcal{U}[0,\, 50]$. With the above settings, we obtain the results  summarized in figures~\ref{fig:2d_gauss} and~\ref{fig:gauss_temp_evolution}. In figure~\ref{fig:2d_gauss_hist} we plot the most probable number of Gaussian pulses present in the data, or in other words, the most probable model given this particular data-set. It is clear that for the given level of noise, it is straightforward to recover the true number of signals. The noise variance is also estimated accurately as $\sigma_n = 0.2\pm2\times10^{-3}$. In figure~\ref{fig:2d_gauss_posteriors} we plot the posterior densities for the parameters of all pulses recovered, while we also mark the true injected values. Figure~\ref{fig:2d_gauss_posteriors} shows the trans-dimensional MCMC chains ``stacked'' over all samples of both model order and model-parameters. As already mentioned, in this simplified scenario all signals have similar value for the amplitude, thus the almost uni-modal marginal on $A_k$. This illustrative example is useful as an introductory application to the more complicated case of detection in Gravitational Wave astronomy presented below, in section~\ref{sec:gws}.

In figure~\ref{fig:gauss_temp_evolution}, three diagnostic quantities for this run are shown. In the top panel, the evolution of temperatures is presented. Following the recipe of ~\citet{Vousden2016}, we control the distances between each temperature chain based on their in-between swap acceptance rate, computed from eq.~(\ref{eq:pt_dt}). The tuning term $\kappa(t)$ is set to $\kappa(t) = t_0/\left(\nu (t + t_0)\right)$, with the {\it adaptation lag} $t_0=10^4$ and the {\it adaptation time} $\nu=10^2$. The middle panel shows the evolution of the swap acceptance rate per number of walkers between the chains running at different temperatures. After $\sim10^5$ sampler iterations, the system converges to an equilibrium, where the rate of swapping states reaches a single value across the temperature range. In the bottom panel we show the acceptance rate for the in-model step of the algorithm, for all temperatures. As expected, after temperature equilibrium at $\sim10^5$ samples, the acceptance rate converges to a (different) value for each temperature, which is higher for higher temperatures (smoother posterior surfaces are easier to explore). 

\begin{figure}
		\centering
		\includegraphics[width=.8\linewidth]{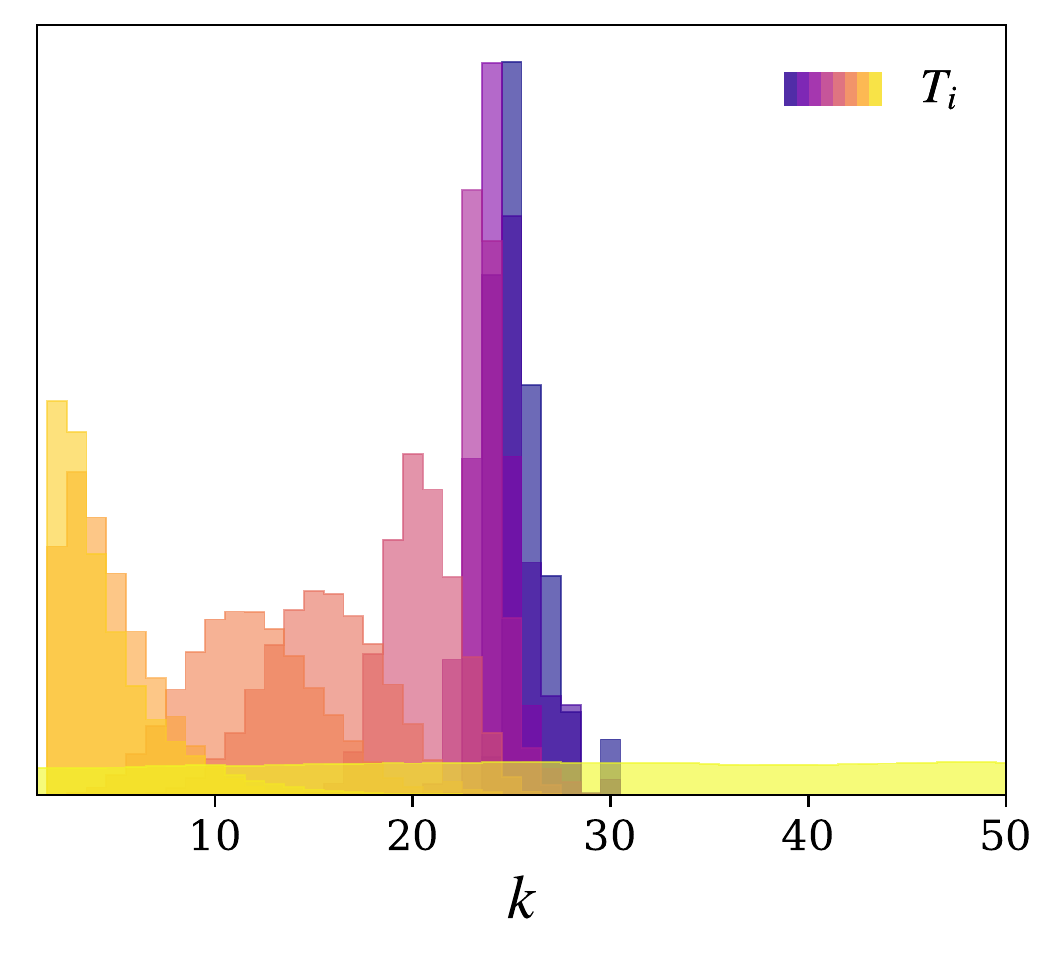}
		\caption{Posterior on the number of Gaussians, $k$, at each temperature $T_i$, for the toy problem of section~\ref{sec:gauss_free}. The different colors indicate the the initial temperature chain index. Darker colors corresponds to colder chains and vice versa.}\label{fig:2d_gauss_k_vs_temp}	
\end{figure}

Finally, it is interesting to investigate how the sampled dimensionality of the problem varies at at different temperatures. In figure~\ref{fig:2d_gauss_k_vs_temp}, we plot the posterior on the number of pulses at each temperature. As expected, higher temperature chains tend to favour lower model dimensionality and the $T_\infty$ chain samples the prior on $k$. This can be attributed to the choice of priors and ``birth proposal'' distributions for both the signal parameters and $k$. The likelihood is down-weighted at higher temperatures, making it harder to overcome the Occam penalty from including extra parameters in the model. This means quieter sources are less likely to be added and the preferred models have fewer sources. 

\subsubsection{\label{sec:psd_splines} Modelling power spectra: Searching for the optimal number of B-spline knots. }

One of the most common problems in signal processing is the characterization of the spectra of the data. This is often done by adopting spectral models and fitting the spectra directly in the frequency domain. This methodology is used when the signal of interest has stochastic properties. Examples from GW astronomy, include the measurement of stochastic signals with astrophysical, or cosmological origin~\citep{lvk_stoch,LISACosmologyWorkingGroup}. There are many examples of possible stochastic signals for 
LISA~\citep{lisa1,lisa2,LISACosmologyWorkingGroup,LISAAstroWorkingGroup}. Searching for signals with stochastic properties requires flexible spectral models, both for the observatory instrumental noise, and the measured stochastic signal. For these reasons, it is sometimes convenient to adopt a versatile model, such as one that is based on B-spline interpolation schemes. 

B-splines are a geometrical modeling tool, and have proven to be very useful for modelling or generating smoother representations of data. They are piece wise polynomial curves with a certain number of continuous derivatives, and can be parameterised in various ways~\citep{PiegTill96}. For this application, we follow~\cite{backgrounders}, and choose to work with cubic-spline interpolation, using the corresponding \textsc{SciPy} library~\citep{virtanen_scipy_2020}. The procedure starts by selecting a number of control points, or {\it knots}, with a given position and amplitude, which the smooth polynomial curve crosses and at which there is a change in the first non-continuous derivative. One of the challenging problems using such methods, is to choose the optimal number of spline knots for fitting the data, without overfitting. This is a model selection problem that can be easily solved with dynamical algorithms such as the one presented here. 

\begin{figure*}	
	\centering
	\begin{subfigure}[t]{2.3in}
		\centering
		\includegraphics[width=1.1\linewidth]{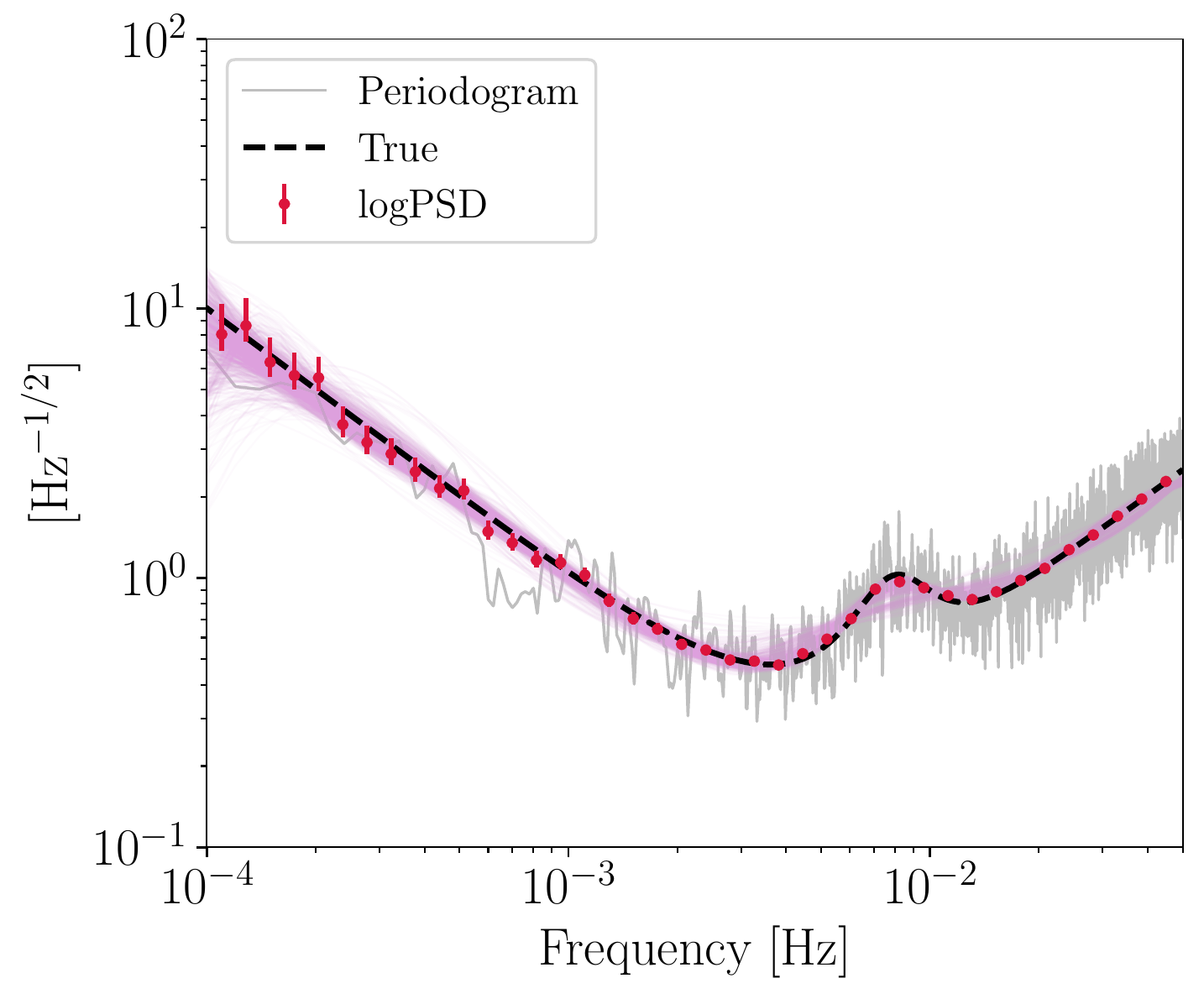}
		\caption{}\label{fig:splinepsddata}		
	\end{subfigure}
	\hspace{0.3em}
	\begin{subfigure}[t]{2.35in}
		\centering
		\includegraphics[width=.86\linewidth]{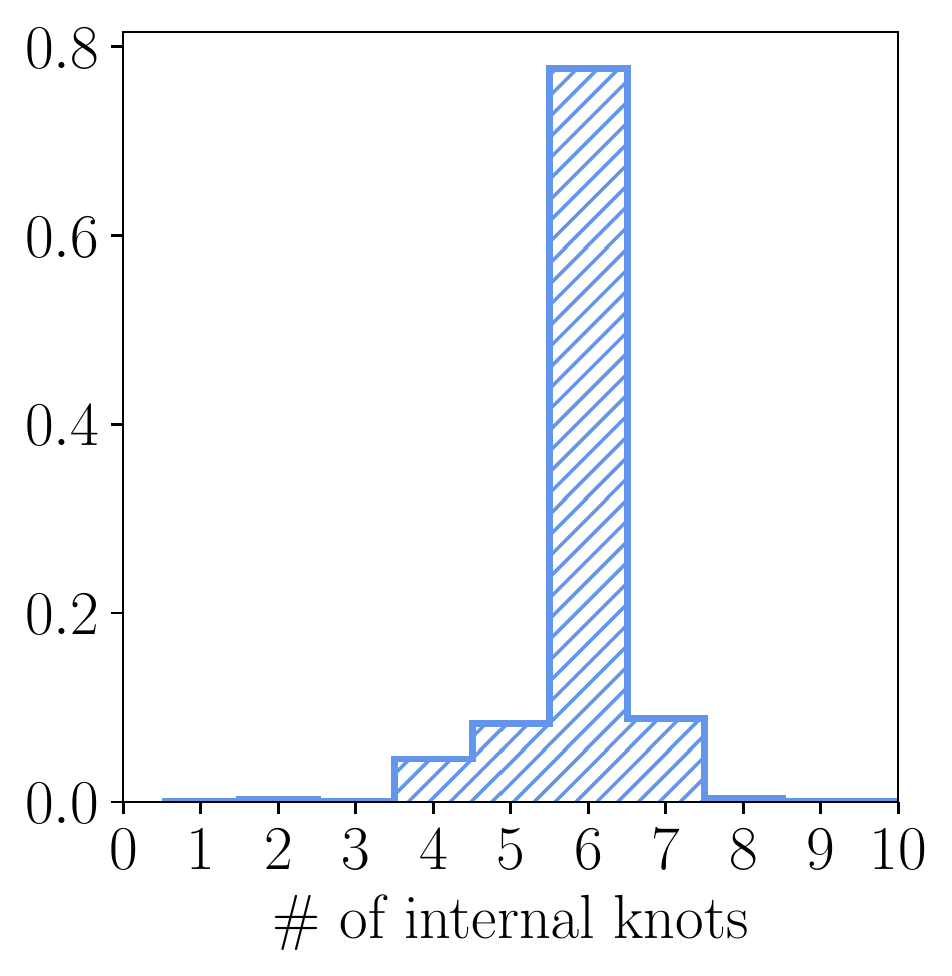}
		\caption{}\label{fig:splinepsdshist}
	\end{subfigure}
	\hspace{-1.8em}
	\begin{subfigure}[t]{2.3in}
		\centering
		\includegraphics[width=.91\linewidth]{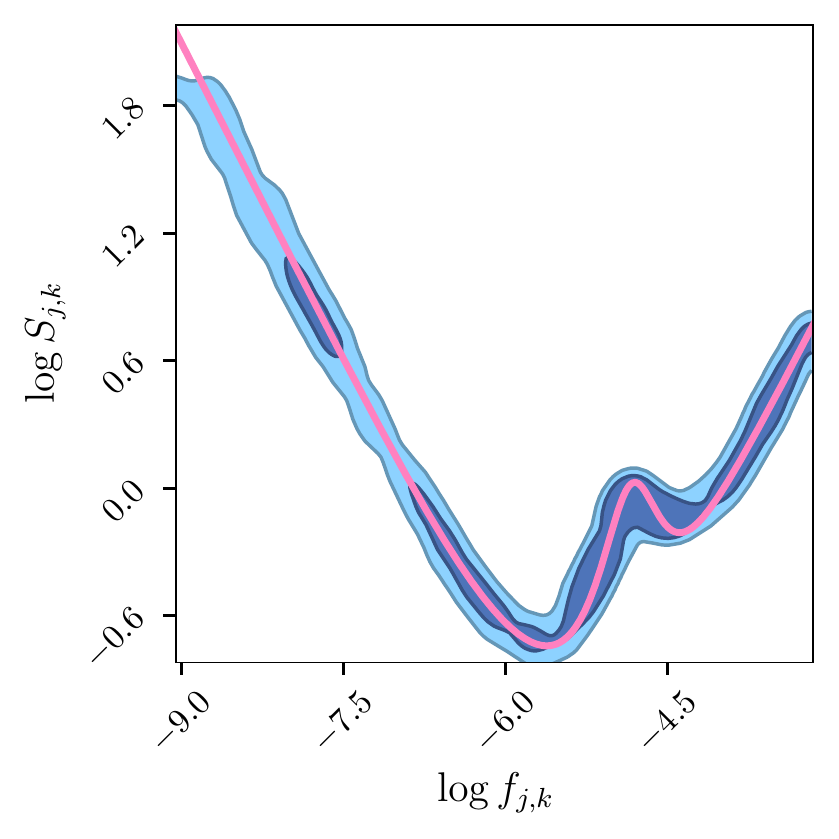}
		\caption{}\label{fig:splinepsdpost}
	\end{subfigure}
	\caption{ Results for power spectra modelling with a shape agnostic model.  (a) The simulated data (gray), generated from the theoretical model (dashed black line). The PSD computed on an equally-spaced logarithmic grid with the method of~\citep{trobsheinzel2006, lpf}, which is used for inference, is represented with the red data points. The pink solid lines represent models drawn randomly from the posterior chains. (b) The optimal B-spline knots estimated by the dynamical parameter estimation procedure. As shown from this histogram, the optimal interior knot count for this data converges to six, corresponding to eight total knots including the two edge knots. (c) Posterior slice for the knot parameters, $(\log f_{j, k},\,\log S_{j, k})$, after stacking the MCMC chains across all model dimensions, $k$. This illustrates where the model prefers to place spline knots, which clearly corresponds to where the spectral density is changing most rapidly. It is also evident from this figure that we essentially ``scan'' the true noise shape (pink solid line), by placing knots across the frequency range (see text for more details). }\label{fig:splinepsd}
\end{figure*}

For our next example, we generate time-series data directly from a theoretical model PSD. The simulated data are represented with the solid gray line in figure~\ref{fig:splinepsddata}. We then use the machinery of \eryn{} to find the optimal dimensionality for the problem, together with the best-fit parameters for the knots. To ease the computational complexity, we compute the PSD of the time series using the methodology developed by~\citet{trobsheinzel2006}. In more detail, we begin by choosing a new frequency grid, on which we compute the PSD using the {\it optimal} number of averaged segments for each given frequency. Following this method, we essentially split the time-series data at the maximum number of segments that the given choice of window and percentage of data overlap permits, which will allow us to estimate the PSD at each frequency bin with minimal variance. By carefully choosing the window function and distance between the data points, one can compute a power spectrum with minimal correlations between frequencies. The estimated spectrum, $D_i$, at each frequency, $f_i$, is then used in the likelihood function given in Eq.~(\ref{eq:spline_likelihood}) below (the spectrum $D_i$ is represented by the red data-points in figure~\ref{fig:splinepsd}). For more details about this method of computing the PSD, we refer the reader to~\citep{lpf, trobsheinzel2006} and for a similar application to the work of~\cite{backgrounders}. 

Finally, we also keep two knots fixed at the edges of the spectra, allowing the sampler to estimate only their amplitude, while the rest of the knot parameters (and their number) are left to be estimated from the data.
For the spline knot positions, $\{\log f_{j, k}\}$, and amplitudes, $\{\log S_{j, k}\}$, we adopt uniform priors that cover the complete parameter space. Here, the $j$ index corresponds to the knot number for the given model order $k$. We also use a ladder of $10$ temperatures, with $10$ walkers each, while maintaining the same settings for the adaptivity of the temperatures as in section~\ref{sec:gauss_free}. Each walker is initialized at a random point on the parameter space, after drawing the dimensionality $k$ of the model from $k\sim\mathcal{U}[1, 20]$. We adopt a Gaussian likelihood, with its logarithm written as
\begin{equation}
	\log p (D | \param_k) \propto -\frac{1}{2}\sum_i n_i \left( \frac{D_i}{{\cal N}_{i,k}(\param_k)} + \log {\cal N}_{i,k}(\param_k)\right),
	\label{eq:spline_likelihood}
\end{equation}
where $D_i$ is the PSD data value for the given frequency $f_i$, as computed by the method presented in~\citep{trobsheinzel2006, lpf}, using $n_i$ averaged segments. The ${\cal N}_{i,k}(\param_k)$ is the spline noise model of order $k$ evaluated at $f_i$, that depends on a parameter set
\begin{equation}
    \param_k = \{\log f_{1,k}, \cdots, \log f_{k,k}, \, \log S_{0}, \cdots, \log S_{k,k},\, \log S_{k+1}\},
    \label{eq:spline_params}
\end{equation}
in which the $\log S_{0}$ and $\log S_{k+1}$ parameters refer to the logarithm of the PSD amplitude of the two fixed knots at the ``edges'' of the spectrum. Those two parameters correspond to our zeroth model order ($k=0$), thus they are always being explored by the walkers of \eryn{}.

The results are shown in figure~\ref{fig:splinepsd}. In particular, in figure~\ref{fig:splinepsdshist} we show the histogram of the recovered number of knots for the particular data-set. It is clear that $8$ spline knots are preferred, two of them being fixed at the edges of the spectrum, and the other six knots free to take any position in the given frequency range. In figure~\ref{fig:splinepsdpost} we show the 2D sliced posteriors for the spline parameters, $\{\log S_{j,k}\}$ and $\{\log f_{j,k}\}$. In this figure, we again stack all the MCMC samples across model orders.  
The true spectrum is indicated by the orange solid line. There is an interesting outcome of this toy investigation; while there is a preferable dimensionality of the model, there is a weak constraint on the actual positions of the knots. We find that the sampler is virtually ``scanning'' the PSD data, showing slightly higher preference for locations between $-6$ and $-4$ in $\log$-frequency, where the spectrum follows a more complicated shape. Finally, in figure~\ref{fig:splinepsddata}, the data (gray solid line and red data points), is shown together with model evaluations drawn from the posterior samples (pink solid lines).

\section{ Examples from Gravitational Wave astronomy}
\label{sec:gws}

In recent years, we have witnessed the beginning of Gravitational Wave Astronomy. Since the first detection~\citep{firstgw} dozens of waveform signatures have been measured with the current network of observatories. At the time of the writing of this paper, more than 90 events have been recorded~\citep{LIGOScientific2021djp}, the vast majority of them are black hole (BH) binary mergers, with a few of them being binary neutron star (NS) and BH-NS mergers. At the same time, detector networks are being improved~\citep{aligo1,abbott2020prospects} and there are plans to expand them with the addition of new observatories, such as the Einstein Telescope~\citep{et1,et2} or Cosmic Explorer~\citep{ce1, abbott2017exploring}. Those detectors will unlock the sky to larger redshifts $z$, allowing access to a vast number of potential sources. In addition, space missions, such as LISA~\citep{lisa1,lisa2}, are predicted to be signal-dominated observatories, with many types of sources populating their data streams. In fact, we expect that source confusion will be one of the primary challenges in future data analysis efforts  in gravitational wave astronomy. In a typical data-set, we expect an unknown number of signals, originating from sources that generate waveforms with different characteristics. Those range from the stellar-mass BH binaries now frequently observed by ground-basd detectors, to the supermassive BH binaries, extreme mass ratio inspirals, ultra compact Galactic binaries (UCB), and stochastic GW signals from both astrophysical and possibly cosmological origin~\citep{lisa1, lisa2, LISACosmologyWorkingGroup}. For this final example, we will focus on the LISA mission, and in particular on the case of discriminating UCB signals. 

\subsection{\label{sec:gbs} Application to LISA data and the Ultra Compact Galactic Binaries}

LISA is going to measure GW signals in the $\mathrm{mHz}$ regime, accessing sources of all the aforementioned types. As already discussed, the most numerous of them are going to be the UCBs, which will be almost monochromatic in the LISA band. Out of the millions of sources,  only $\sim\mathcal{O}(10^4)$ will be individually resolvable, and the rest will generate a confusion signal. As a consequence, for the duration of the mission, we will need to disentangle tens of thousands of sources which will be overlapping in both time and frequency domains. This is no trivial task, but various different strategies have already been proposed for analyzing such challenging data-sets. For example, Gaussian Processes can be employed~\citep{Strub2022upl}, or Swarm Optimization techniques~\citep{Zhang2021}, or hybrid swarm-based algorithms~\citep{Bouffanais2016}. Pipelines based on MCMC methods have been tested extensively~\citep{crowder2007, Littenberg2011, rjmcmc_app0, protoglobal}, and have been demonstrated to be able to tackle complex cases where signals are overlapping. 

\begin{figure}
 	\includegraphics[width=1\linewidth]{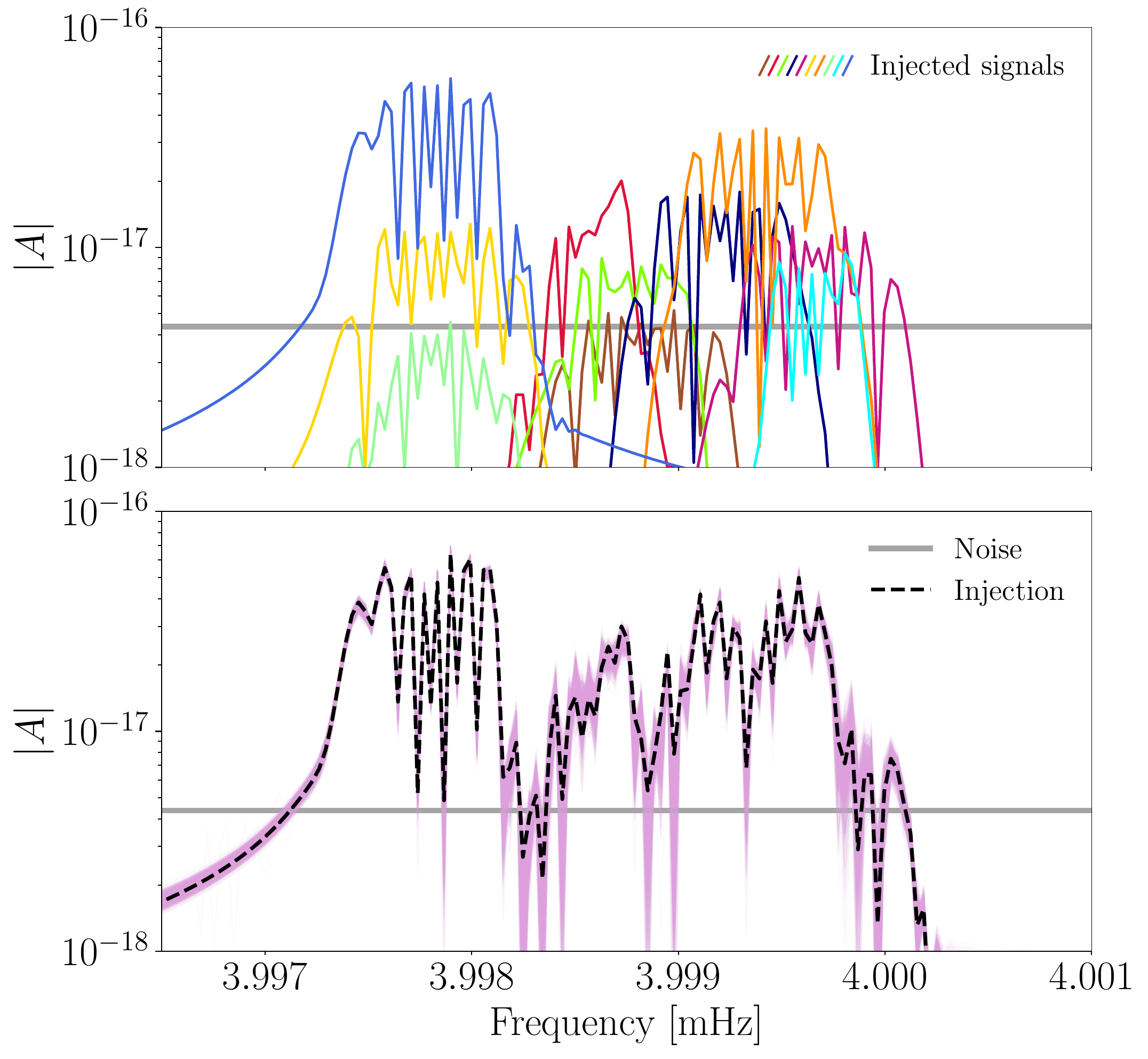}
 	\caption{ {\it Top panel}: The simulated data used for demonstrating the capabilities of \eryn{} in tackling a high-dimensional problem. A total of 10 Ultra Compact Binaries in the vicinity of our Galaxy emitting Gravitational Wave signals at the $\mathrm{mHz}$ range. Here we plot the power spectral density of the $A$ data channel of LISA. The catalogue of sources is taken from the second LISA Data Challenge~\citep{ldc}. Each signal is represented by a different colour. {\it Bottom panel}: The same data-set, now comparing the injected signal against the solution yielded by \eryn{}~(see text for more details). We have plotted the shaded area by sampling the joint posterior on model order $k$ and the corresponding parameters. }
\label{fig:gbs_residuals}
\end{figure}

Here, we will focus on the same problem, employing \eryn{} to solve a down-scaled version of the UCB challenge. It is down-scaled because we focus only on a single narrow frequency band, containing several overlapping signals, in the presence of instrumental noise only\footnote{No confusion signal from other unresolved UCBs is considered in this investigation.}. In addition, we focus solely on demonstrating the capabilities of \eryn{} on dynamic parameter estimation for UCB type sources and no other types of signals are contained in the data (e.g., chirping signatures from supermassive BH binaries). At the same time, we have access to the level of instrumental noise, which is shown in both panels of figure~\ref{fig:gbs_residuals}. Searching for the UCB signals across the complete LISA band requires a more elaborate implementation of this simplified pipeline. This pipeline will be focusing on solving the complete second LISA Data Challenge~\citep{ldc}, and is going to be presented in future work. 
\begin{table}
\centering
\begin{tabular}{|c|  p{2cm} p{1cm} |} 
 \hline 
 \# & \centering $f_\mathrm{gw}~[\mathrm{mHz}]$ & $\rho_\mathrm{opt}$ \\ [1ex] 
 \hline\hline
 1  & \centering 3.99780 & 9.98  \\ [1ex] 
 2  & \centering 3.99781 & 46.70  \\ [1ex] 
 3  & \centering 3.99784 & 4.55  \\ [1ex] 
 4  & \centering 3.99854 & 39.45  \\ [1ex] 
 5  & \centering 3.99873 & 13.02  \\ [1ex] 
 6  & \centering 3.99882 & 8.47  \\ [1ex] 
 7  & \centering 3.99919 & 10.88  \\ [1ex] 
 8  & \centering 3.99939 & 19.07  \\ [1ex] 
 9  & \centering 3.99964 & 20.00  \\ [1ex] 
 10 & \centering 3.99965 & 7.99  \\ [1ex] 
 \hline
\end{tabular}
\caption{The optimal SNR $\rho_\mathrm{opt}$ for each of the 10 injected sources, computed for the given duration of the mission (see eq.~(\ref{eq:asnr})). The dominant emission frequency $f_\mathrm{gw}$ is also given for reference.}
\label{table:ucbs}
\end{table}
We choose to work on the frequency segment between $3.997$ and $4~\mathrm{mHz}$, which contains $10$ UCB objects, drawn directly from the LDC2 catalogue~\citep{ldc}. Those are shown in the top panel of figure~\ref{fig:gbs_residuals} which shows the power spectrum of the $A$ data channel of LISA. We use the two noise-orthogonal $A$, $E$, and $T$ Time Delay Interferometry variables~\citep{tdi, aet, baghi2021}, which are linear combinations of the LISA relative frequency TDI Michelson measurements $X$, $Y$, and $Z$ as:
\begin{equation}
\begin{aligned}
    A &= \frac{1}{\sqrt{2}}(Z - X), \quad E = \frac{1}{\sqrt{6}}(X - 2Y + Z), \\
    T &= \frac{1}{\sqrt{3}}(X + Y + Z).
    \label{eq:aet}
\end{aligned}
\end{equation}
In ideal conditions (equal noises across spacecrafts, and equal LISA arms), the noise in $A$ and $E$ is independent, while the $T$ data stream can be used as a signal-insensitive {\it null} channel, useful for instrument noise calibration. Since we perform analysis on a noise-free injection, we will be neglecting the $T$ channel altogether. We simulated the injection data for an observation time of $\mathrm{T}_\mathrm{obs} = 1~\mathrm{year}$. 

The {\it optimal} SNR for each injected source, $\rho_\mathrm{opt}$, is given in table~\ref{table:ucbs}. The $\rho_\mathrm{opt}$ quantity refers to the SNR of each source in {\it isolation}, with respect only to the instrumental noise, and can be calculated as
\begin{equation}
    \rho_\mathrm{opt}^2 = \sum_C  \left( h_C | h_C \right)_C,
    \label{eq:snrtot}
\end{equation}
with $C \in \{A, E \}$ the noise-orthogonal TDI channels of eq.~(\ref{eq:aet}), while the $\left( \cdot | \cdot \right)$ notation represents the noise weighted inner product expressed for two time series $a$ and $b$ as
\begin{equation}
\left( a | b \right) = 2 \int\limits_0^\infty \mathrm{d}f \left[ \tilde{a}^\ast(f) \tilde{b}(f) + \tilde{a}(f) \tilde{b}^\ast(f) \right]/\tilde{S}_{n}(f).
\label{eq:ineerprod} 
\end{equation}
The tilde represents the data in the Fourier frequency domain, and the asterisk indicates complex conjugate.
The $\tilde{S}_{n}(f)$ is the one-sided PSD of the noise for a given TDI channel. Under our assumptions $S_{n,A}(f)=S_{n,E}(f)$.

For our investigation we chose to analyze noiseless data (no noise realization), while in the likelihood we are using the PSD noise levels taken from the LISA design studies~\citep{scird}. For the signals, we utilize the fast frequency-domain UCB waveform model of~\citet{rjmcmc_app1}. Then, the two polarizations of an emitting UCB can be written as
\begin{equation}
\begin{aligned} 	
    h_+(t)& = \frac{2\mathcal{M}}{D_L}\left(\pi f_\mathrm{gw}(t)\right)^{2/3}\left( 1 + \cos^2 \iota \right) \cos{\phi(t)}, \\
	h_\times(t)& = -\frac{4\mathcal{M}}{D_L}\left(\pi f_\mathrm{gw}(t)\right)^{2/3}\cos{\iota} \sin{\phi(t)}, \label{eq:ucbhphc}
\end{aligned}
\end{equation}
where $\mathcal{M}$ is the chirp mass, $f_\mathrm{gw}$ is the instantaneous gravitational wave frequency, $D_L$ is the luminosity distance, $\iota$ is the inclination of the binary orbit, and $\phi(t)$ is the gravitational wave phase. The phase $\phi$ can be expressed as $\phi=\phi_0 + 2\pi \int^t f_\mathrm{gw}(t')dt'$, with $\phi_0$ being an initial arbitrary phase shift. The LISA constellation is assumed to be rigid and with equal arms, while the spacecrafts are assumed to follow analytic Keplerian orbits~\citep{PhysRevD.67.022001}. Under these assumptions, it is straightforward to compute its response to the almost monochromatic waveforms of eqs.~(\ref{eq:ucbhphc})~\citep{PhysRevD.67.022001, Babak:2021mhe}. For more details about the waveform model, the response of the instrument, and the orbits of the constellation, we refer the reader to~\citet{rjmcmc_app1, robson2018, katz2022, PhysRevD.67.022001, Babak:2021mhe}. 

In our simplified scenario, each binary signal in the Solar System Barycenter is determined by a set of eight parameters. Those are the $\param = \{ \mathcal{A},\, f_\mathrm{gw}~[\mathrm{mHz}],\, \dot{f}_\mathrm{gw}~[\mathrm{Hz/s}],\, \phi_0,\, \cos\iota, \psi, \lambda, \sin\beta \}$, where $\mathcal{A}$ is the overall amplitude, $\dot{f}_\mathrm{gw}$ is the first derivative of the gravitational-wave frequency, $\psi$ the polarization, $\lambda$ is the ecliptic longitude, and $\beta$ the ecliptic latitude of the binary. The amplitude of the signal is calculated as
\begin{equation}
	\mathcal{A} = \left(2 \mathcal{M}^{5/3}\pi^{2/3} f_\mathrm{gw}^{2/3} \right) / D_L, 
	\label{eq:amplitude}
\end{equation}
which can be used to obtain a rough SNR estimate, via~\citep{rjmcmc_app0}
\begin{equation}
	\rho^2 = \frac{\mathcal{A}^2 \mathrm{T}_\mathrm{obs}\sin^2(f_\mathrm{gw}/f_\ast)}{4 S_n(f_\mathrm{gw})}, 
	\label{eq:asnr}
\end{equation}
with $S_n(f_\mathrm{gw})$ being the instrumental noise power spectral density at frequency $f_\mathrm{gw}$, and $f_\ast = 1 / (2 \pi L)$, where $L$ the LISA arm length. Given eq.~(\ref{eq:amplitude}) and~(\ref{eq:asnr}), we find it convenient to directly sample on $\rho$ instead of $\mathcal{A}$, which also yields a more illustrative measure of the amplitude of each binary. Then, if $d$ is the measured TDI data and $\mathrm{h}$ the given GW signal after applying the response of the instrument, the logarithm of the likelihood for an arbitrary number $k$ of UCB signals can be written as
\begin{equation}
        \log p (d | \param_k) \propto \left(d | \mathrm{h}_k \right) - \frac{1}{2} \left( \mathrm{h}_k | \mathrm{h}_k \right), 
	\label{eq:ucbllh}
\end{equation}
where for the sake of convenience, we have defined $\mathrm{h}_k = \sum_k \mathrm{h}(\param_k)$.

We use wide uniform priors for all the rest of the binary parameters, covering essentially the complete parameter space. The exception is again the amplitude (SNR), $\rho$, where we adopt a prior which was first introduced in~\citet{Cornish2015} and then adapted in~\citet{rjmcmc_app0}. The prior density can be expressed as
\begin{equation}
	p(\rho) = \frac{3\rho}{4\rho_\ast \left(1 + \rho / (4\rho_\ast) \right)},
	\label{eq:aprior}
\end{equation}
where $\rho_\ast$ is a given constant that specifies the peak of the above distribution. This distribution is designed to prevent the proposal of sources with very small SNR in the model, as it drops sharply for $\rho\rightarrow0$. Those weak sources do not significantly affect the likelihood, and so their inclusion must be penalised by the prior~\footnote{We remind that the prior is also used as our ``birth'' proposal here.}. This prior choice forces the sampler to explore only potential sources with non-zero SNR, avoiding populating the chains with numerous undetectable signals. This prior performs adequately in this problem, but there are other solutions one could adopt in order to keep control of the number of very weak sources. This discussion, which sets the grounds for a {\it global-fit} analysis pipeline for the LISA data~\citep{rjmcmc_app0}, is out of the scope of this paper, but a more detailed description will be presented in a future work. 
\begin{figure*}	
	\centering
	\begin{subfigure}[t]{3in}
		\centering
		\includegraphics[width=.75\linewidth]{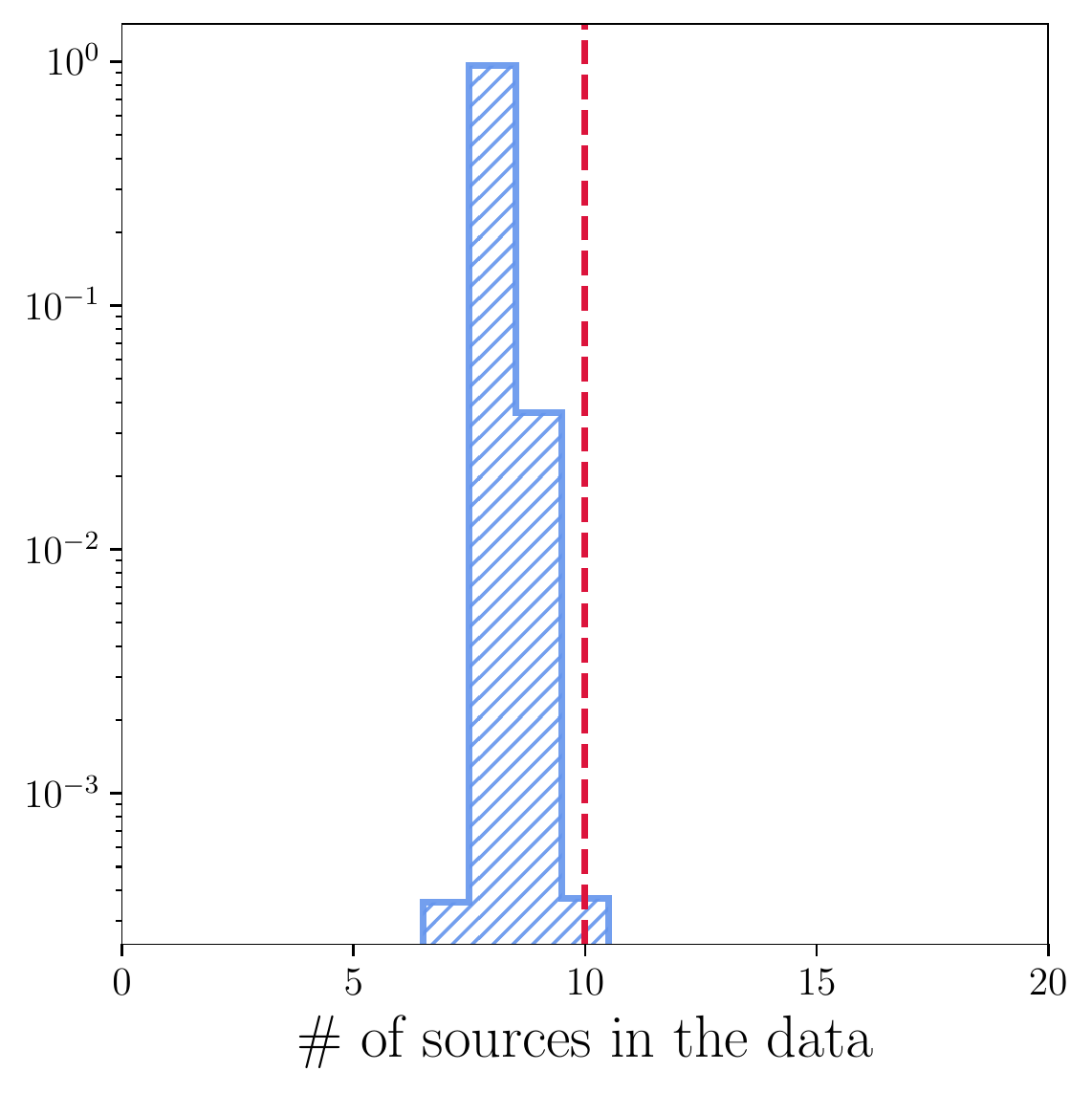}
		\caption{}\label{fig:ucbs_source_hist}		
	\end{subfigure}
	\quad
	\begin{subfigure}[t]{3in}
		\centering
		\includegraphics[width=.8\linewidth]{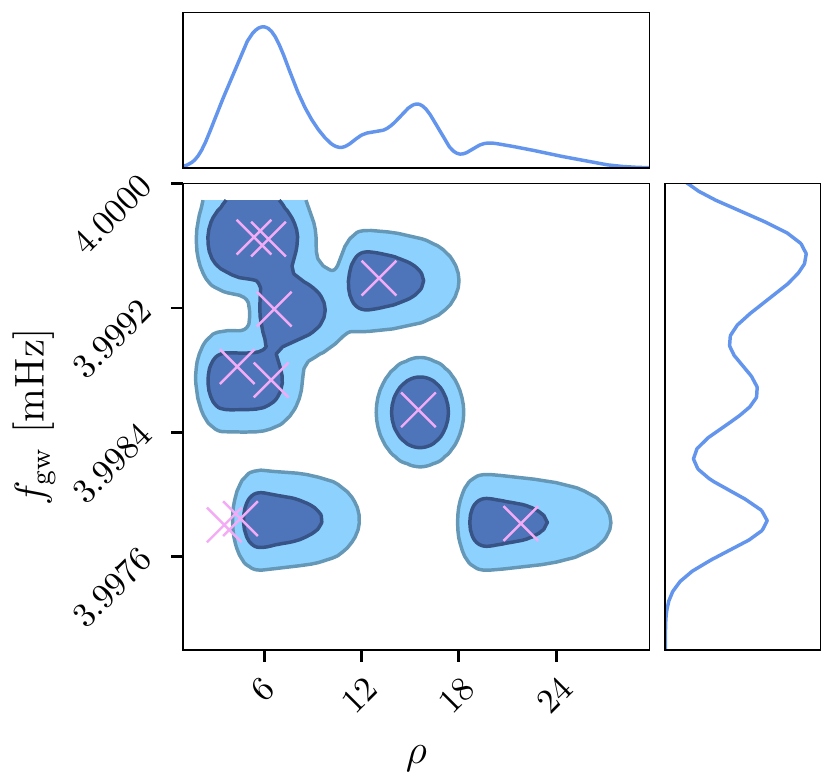}
		\caption{}\label{fig:ucbs_a_f0}
	\end{subfigure}
	\caption{{\it Left panel}: In this figure, we show the posterior on the number of UCB sources in the data. The true injected number is shown with the red dashed line. It is clear that, for the given measurement duration of the particular data set, we manage to confidently resolve eight binaries out of a total of ten. {\it Right panel}: Corner plot for two of the eight parameters characterising each UCB source. These are the amplitude, expressed as an SNR $\rho$, and the dominant emission (or initial) frequency, $f_\mathrm{gw}~[\mathrm{mHz}]$ (see text for more details). The violet crosses represent the injected parameter values. A corner plot for more parameters is shown in figure~\ref{fig:ucbs_triangle_more_params} in the Appendix.}
\label{fig:ucbs}
\end{figure*}
\paragraph{Search Phase:} Before parameter estimation, we initiate a {\it search phase} of our analysis, with the aim of getting the walkers to a better starting point on the posterior surface. This phase consists of an iterative brute force procedure, based on drawing a very large number of proposals, then maximizing the likelihood over the initial phase $\phi_0$, and finally perform a rapid MCMC sampling over the parameter space, using only a one-source model (therefore there is no dynamical parameter spaces). In particular, we draw $5\times10^6$ points in the parameter space, and after phase maximization, we use them as starting points to a parallel tempered MCMC run with $N_T = 10$ temperatures, each running with $n_w=500$ walkers. When this step concludes, we keep the 100 best samples in terms of likelihood value and use their corresponding parameter estimates as starting points for the parameter estimation portion of the analysis. We also use the maximum Likelihood solution to subtract the source found from the data. We then use the residual data to search for another source, and this process repeats until there is no signal found with SNR $\rho>5$. In between successive iterations of the single-binary search, we run another MCMC over all sources found so far in order to adjust the parameters to account for correlations and overlap between sources. After convergence, we found eight sources in our data-set with an optimal SNR greater than 5. We take these found sources and add them to all walkers in the sampler at the beginning of the full MCMC run described below.

\paragraph{Parameter estimation:} During this step, we perform hybrid MCMC sampling, where we both update the found sources (in-model) and dynamically search for new and weaker sources in the data employing Reversible Jump sampling. For the number of signals $k$, we adopt a uniform prior $k\sim\mathcal{U}[6, \, 20]$. For the sake of convenience in this simple application, we keep the six loudest binaries found during the search phase as {\it fixed}. This means that we still sample their waveform parameters, but they are not allowed to be removed by the Reversible Jump process. We chose this setup in order to accelerate the convergence of the algorithm, being confident that these sources are part of the solution. In future work, this will be adjusted to deal with the much larger complexity of the full problem. 

Concerning the sampler settings, we use the adaptive parallel tempering scheme of section~\ref{sec:pt}, building a temperature ladder of $N_T = 10$ temperatures, with $100$ walkers for each temperature. For this run, we have also utilized the Multiple Try Metropolis algorithm (see section~\ref{sec:mt}) in order to improve the acceptance rate in the reversible jump proposal. We have also tried the Delayed-Rejection scheme which is implemented for \eryn{}, but we found that the Multiple-Try strategy yields more efficient sampling. Finally, we have utilized the basic stretch and group stretch proposals that were described in section~\ref{sec:emcee}.

After convergence, the result is shown in figures~\ref{fig:gbs_residuals} and~\ref{fig:ucbs}. In figure~\ref{fig:ucbs_source_hist}, the sampled posterior on the number of sources $k$ is presented. In this histogram, we have added the six fixed binaries to the actual number of signals being sampled via the Reversible Jump algorithm. It is fairly obvious that we have managed to confidently resolve eight out of the ten injected binary signals. This fact that we do not favour $10$ sources can be explained partly by the low SNR of the signals (see table~\ref{table:ucbs}) and partly by confusion from source overlap (also shown in figure~\ref{fig:gbs_residuals}). Additionally, the result of figure~\ref{fig:ucbs_source_hist} depends on the given observation duration. The greater the $T_\mathrm{obs}$, the better our ability to resolve the confused sources. Thus, in that case, we should expect more Reversible Jump iterations across the higher dimensional models.

On the right panel, in figure~\ref{fig:ucbs_a_f0}, the ensemble 2D posterior slice is shown, for two selected parameters. We call it ensemble because we are again ``stacking'' all the chains for these two parameters for all sources for all model orders $k$. We chose to show only the amplitude (the $\rho$ parameter explained in eq.~(\ref{eq:asnr}) above) and the dominant emission frequency $f_\mathrm{gw}$, which illustrates the number of sources resolved, and how they overlap in frequency. A corner plot for more parameters is shown in figure~\ref{fig:ucbs_triangle_more_params} in Appendix~\ref{sec:apppendix}. We also show the true injection values, marked as crosses, on top of the 2D posterior. From this plot alone, one can see that the sampler is exploring efficiently the parameter space, converging to the true values of the resolvable binaries that were injected. 

\section{Discussion}
\label{sec:discussion}

We have implemented \eryn{}, a Bayesian sampling package capable of performing efficient trans-dimensional inference, by employing different techniques that improve its acceptance rate. These techniques are the affine invariant sampling, the adaptive parallel tempering, the delayed rejection, and multiple try metropolis, in combination with the construction of informative proposal distributions for the parameters of the models. The structure of \eryn{} is based on the widely used software \emcee{}~\citep{emcee}, enhanced with the ability of performing Reversible Jumps~\cite{rjmcmc1} between different model spaces. The sampler capabilities have been demonstrated with toy models that are commonly encountered in different data analysis problems. We have begun with an application to signal detection, and in particular to searching for simple signals in the form of Gaussian pulse signals in the presence of Gaussian noise (see section~\ref{sec:gauss_free}). 

In section~\ref{sec:psd_splines}, we applied our algorithm to a problem of modeling power spectra with arbitrary shapes in frequency domain. In such cases, it is convenient to define models based on B-splines, which are able to faithfully capture  the shape of any spectral data. However, in order to avoid over-fitting situations, the optimal order of the model  ({\it i.e.} the optimal number of spline knots), needs to be estimated from the data. This can be done either sequentially, by trying models of different dimensionality and then comparing their performance, or dynamically, by using trans-dimensional algorithms such as \eryn{}. This class of problems is often encountered in cosmology~\citep{planck_2018,planck_spectra}, where the signal of interest is stochastic in nature, and sometimes the prior knowledge on its shape is very limited. As already discussed, this is especially true for future GW observatories, which open the possibility of detecting such signals from both astrophysical and cosmological origin~\citep{lisa1, LISACosmologyWorkingGroup,backgrounders}. The different theoretical models produce spectra with distinct shapes, increasing the need for shape-agnostic spectral models, such as the B-spline used here.

Finally, in section~\ref{sec:gbs}, we demonstrated \eryn{} in a more complicated problem, that of the analysis of ultra compact binary signals measured by the future LISA detector. These objects are going to produce the majority of the signals in the LISA data, each emitting almost monochromatic radiation. Their vast number will generate a confusion foreground, while only a few thousand of them will be resolvable from the data. We employ our tools described in this work, together with a search phase that is based on iteratively running the sampler on ``static models'' (no trans-dimensional moves) with phase-maximized likelihoods. We do these runs on the residuals of each iteration, with the aim of extracting all bright sources. In order to achieve faster convergence of our parameter estimation run, we choose to keep the brightest sources found during search as fixed (minimum number of model order is $k = 6$), while the Reversible Jump algorithm is used to search for weaker signals in the data. This is purely a choice that allows quick convergence in this fully-controlled and simplified LISA data-set.

We perform this analysis for a mission duration of $T_\mathrm{obs}=1~\mathrm{year}$ and only on a single narrow frequency band around~$4~\mathrm{mHz}$, which contains a total of $10$ binary signals. It is worth noting here, that the synthetic data were produced assuming idealized conditions. This means that we do not consider any data irregularities, such as data gaps and glitches and spectral lines, or any other contamination originating from the mixing of signals of different types (such as supermassive BH binaries). In the end, as shown in figure~\ref{fig:ucbs_source_hist}, we manage to recover eight out of ten injected signals. This result makes sense given the relative strength of the injections, and their waveform overlap. Many of the injected sources have an optimal SNR in isolation which is rather low (see table~\ref{table:ucbs}), so these are more susceptible to deterioration when we account for signal overlap.

The above investigations demonstrate that the dynamical parameter estimation capabilities of \eryn{} are suitable for these types of problems. This feature is missing from already existing libraries such as \texttt{bilby}~\cite{bilby}, which are used by the Gravitational Wave community. \texttt{bilby} offers a wide selection of tools which are necessary for the Bayesian analysis of Gravitational Wave data. These include implementations of likelihood and prior functions, instrument response models and spectral densities, waveform models, and a number of samplers to choose from (both MCMC and nested samplers). On the other hand, \eryn{} is based on parallel tempering MCMC enhanced with Reversible Jump, which allows Bayesian analyses for a wider set of problems, in which sampling of a dynamical parameter space is needed.

\eryn{} has already been used in several works that have been already published~\citep{katz2022, 2022PhRvD105d4055K, backgrounders, Sasli:2023mxr}, or are going to appear soon. The work presented in this paper is the initial part of our efforts toward implementing a data analysis pipeline for LISA data. This pipeline will be demonstrated on the LDC2 data-set~\citep{ldc}, which contains multiple types of signals overlapping in both time and frequency domains. That being said, \eryn{} is a generic and versatile sampler, which can be used in any investigation that requires Reversible Jump sampling, and to our knowledge is one of the very few statistical tools of this kind that is not specialized to a single type of analysis (see discussion in section~\ref{sec:rj}).

\section*{Acknowledgements}
We wish to thank S. Babak, M. Le Jeune, S. Marsat, T. Littenberg, and N. Cornish for their useful comments and very fruitful discussions. N Stergioulas and N Karnesis acknowledge  support from the Gr-PRODEX 2019 funding program (PEA 4000132310). 

\section*{Data Availability}
\label{sec:data_availability}
\noindent

The sampler code can be found at \url{https://github.com/mikekatz04/Eryn}. The code for the toy examples of section~\ref{sec:examples} can be found in the documentation of \eryn{} (\url{https://mikekatz04.github.io/Eryn/html/index.html#}). For the analysis of section~\ref{sec:gws}, we used the {\tt GBGPU} package (\url{https://github.com/mikekatz04/GBGPU}), which contains an implementation of the UCB waveforms of the LDC software (\url{https://lisa-ldc.lal.in2p3.fr/code}).

\bibliographystyle{mnras}

\appendix
\renewcommand\thefigure{\thesection.\arabic{figure}} 
\section{}
\label{sec:apppendix}

\begin{figure*}
 	\includegraphics[width=.8\linewidth]{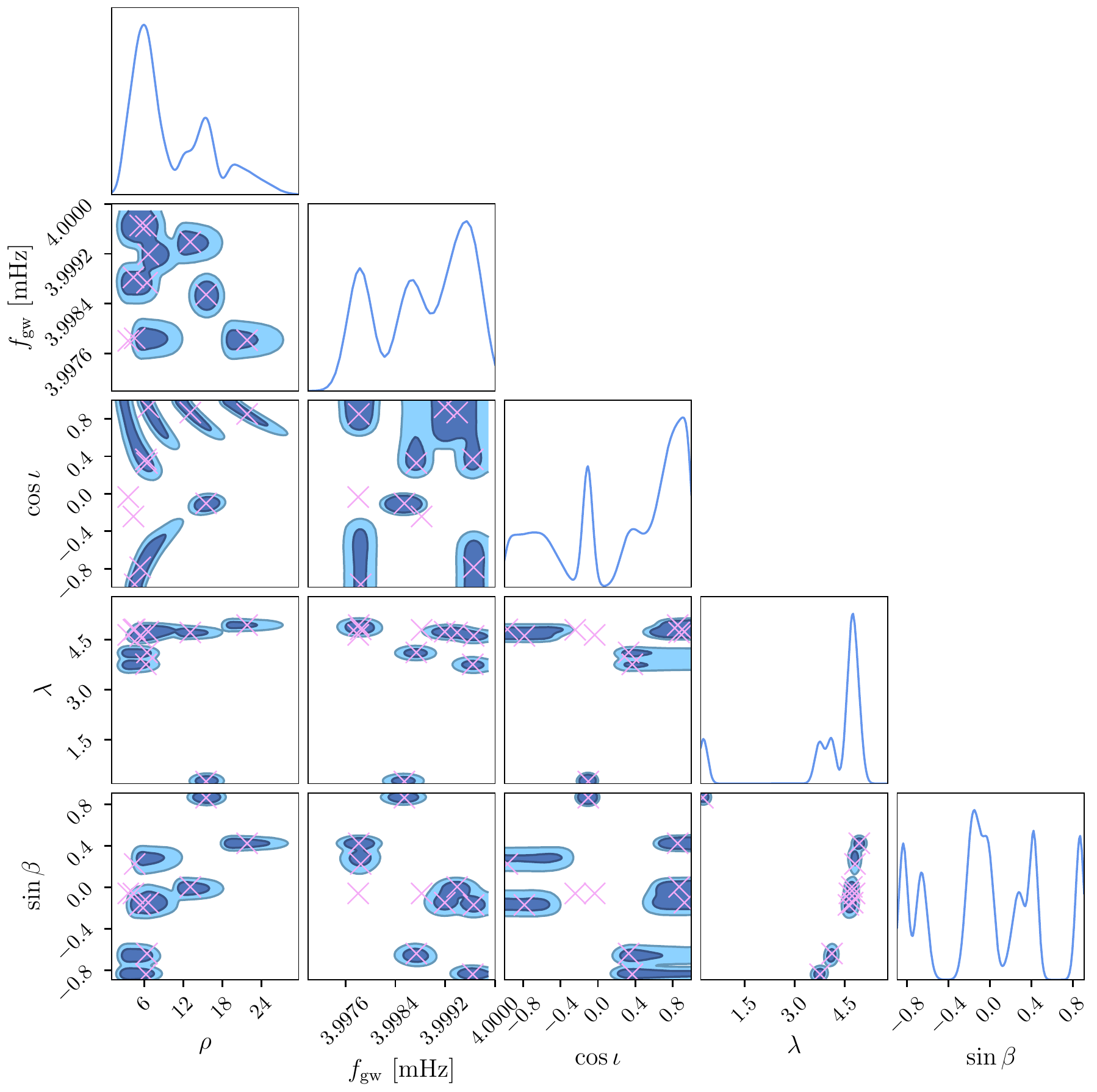}
 	\caption{ A triangle plot showing the 2D posterior slices for the application of section~\ref{sec:gbs}, but for a greater selection of parameters than figure~\ref{fig:ucbs_a_f0}. The rest of the parameters, if plotted stacked in the same manner, result in surfaces that cannot be so easily interpreted, and therefore have been left out. The true injected values are marked with crosses.  
 	\label{fig:ucbs_triangle_more_params} }
\end{figure*}

In figure~\ref{fig:ucbs_triangle_more_params} we show the triangle plot of the stacked posterior points as sampled by \eryn{}, for the investigation of section~\ref{sec:gws}. The difference to figure~\ref{fig:ucbs_a_f0} is that here we include more parameters of the sources, but we still do not include all parameters for the sake of clarity.

\end{document}